\pdfoutput=1

\documentclass[structabstract]{aa}  

\usepackage{graphicx}
\usepackage{txfonts}
\usepackage{natbib}
\usepackage{rotating}
\usepackage{wasysym}

\newcommand{\ergpersec}{erg\,s$^{-1}$}

\newcommand{\ergperseccm}{erg\,s$^{-1}$\,cm$^{-2}$}

\newcommand{\phperseccmkev}{ph\,s$^{-1}$\,cm$^{-2}$\,keV$^{-1}$}
\newcommand{\phperseccm}{ph\,s$^{-1}$\,cm$^{-2}$}
\newcommand{\ctspersec}{counts\,s$^{-1}$}
\newcommand{\cm}{cm$^{-2}$}
\newcommand{\degree}{$^{\circ}$}
\newcommand{\micron}{$\mu$m}
\newcommand{\st}{$^{\rm st}$}
\newcommand{\nd}{$^{\rm nd}$}
\newcommand{\rd}{$^{\rm rd}$}
\newcommand{\ith}{$^{\rm th}$}

\newcommand{\sgra}{Sgr\,A$^\star$} 
\newcommand{\sgraigr}{IGR\,J17456--2901}
\newcommand{\sgrahess}{HESS\,J1745--290}
\newcommand{\sgrafermi}{1FGL\,J1745.6--2900}

\newcommand{\pwin}{G\,359.95--0.04}

\newcommand{\xmm}{\it XMM-Newton\rm}
\newcommand{\xmmshort}{\it XMM\rm}
\newcommand{\integ}{\it INTEGRAL\rm}

\newcommand{\chandra}{\it Chandra\rm}

\newcommand{\apex}{\it APEX\rm}
\newcommand{\vlt}{\it VLT\rm}
\newcommand{\keck}{\it Keck\rm}
\newcommand{\hst}{\it HST\rm}
\newcommand{\subaru}{\it Subaru\rm}
\newcommand{\sma}{\it SMA\rm}
\newcommand{\cso}{\it CSO\rm}
\newcommand{\fermi}{\it Fermi\rm}
\newcommand{\hess}{\it HESS\rm}

\begin{document}
   \title{Concurrent X-ray, near-infrared, sub-millimeter, \\ and GeV gamma-ray observations of \sgra}
   \subtitle{}
   \author
   	{G.~Trap
          \inst{1,2,3}\fnmsep\thanks{\email{trap@apc.univ-paris7.fr}}
         \and
          A.~Goldwurm
          \inst{1,2}
          \and
           K.~Dodds-Eden
          \inst{4}
          \and
          A.~Weiss
         \inst{5}
          \and
          R.~Terrier
          \inst{2}
          \and
          G.~Ponti
	\inst{2,6}
	\and
	 S.~Gillessen
      \inst{4}
          \and
          R.~Genzel
          \inst{4}
          \and
          P.~Ferrando
          \inst{1,2}
          \and
         G.~B\'elanger
          \inst{7}
          \and
          Y.~Cl\'enet
          \inst{8}
          \and
          D.~Rouan
          \inst{8}
          \and
          P.~Predehl
          \inst{4}
          \and
          R.~Capelli
          \inst{4}
          \and
           F.~Melia 
          \inst{9}
          \and
          F.~Yusef-Zadeh
          \inst{10}
                }

   \institute
   	{Service d'Astrophysique (SAp) / IRFU / DSM / CEA-Saclay -- 91191 Gif-sur-Yvette, France
         \and
           AstroParticule \& Cosmologie (APC) / Universit\'e Paris VII / CNRS / CEA / Observatoire de Paris --  75205 Paris, France
           \and
           D\'epartement de Physique / Palais de la d\'ecouverte / Universcience -- 75008 Paris, France
            \and
           Max Planck Institut f\"ur Extraterrestrische Physik (MPE) -- 85748 Garching, Germany
           \and
           Max Planck Institut f\"ur Radioastronomie (MPIfR) -- 53121 Bonn, Germany
           \and
           School of Physics and Astronomy / University of Southhampton -- Highfield, Southhampton, SO17 1BJ, UK
           \and
           European Space Agency (ESA) / ESAC -- P.O. Box 778, Villanueva de la Canada, 28691 Madrid, Spain
           \and
          Laboratoire d'\'etudes spatiales et d'instrumentation en astrophysique (LESIA) / Observatoire de Paris -- 92195 Meudon, France
           \and
           Department of Physics and Steward Observatory / University of Arizona, Tucson, Arizona 85721, USA
       \and
           Department of Physics and Astronomy / Northwestern University, Evanston, Illinois 60208, USA
            }

   \date{Received ... ; accepted ...}
  \abstract
     {}
   {The radiative counterpart of the supermassive black hole at the Galactic center (GC), \sgra, is subject to frequent flares visible simultaneously in X-rays and near-infrared (NIR). Often, enhanced radio variability from centimeter to sub-millimeter wavelengths is observed to follow these X-ray/NIR eruptions. We present here a multi-wavelength campaign carried out in April 2009, with the aim of characterizing this broadband flaring activity.}
   {Concurrent data from the \xmm/EPIC (2--10~keV), \vlt/NACO (2.1~\micron, 3.8~\micron), \apex/LABOCA (870~\micron), and \fermi/LAT (0.1--200~GeV) instruments are employed to derive light curves and spectral energy distributions of new flares from \sgra.}
   {We detected two relatively bright NIR flares both associated with weak X-ray activity, one of which was followed by a strong sub-mm outburst $\sim$200~min later. Photometric spectral information on a NIR flare was obtained for the first time with NACO giving a power-law photon index $\alpha=-0.4\pm0.3$ ($F_{\nu}\propto \nu^{\,\alpha}$). The first attempt to detect flaring activity from the \fermi\ GC source \sgrafermi\ is also reported. NIR, X-ray, and sub-mm flares are finally modeled in the context of non-thermal emission processes. It is found that the simplest scenario involving a single expanding plasmoid releasing synchrotron NIR/sub-mm and synchrotron self-Compton X-ray radiation is inadequate to reproduce the data, but suggestions to reconcile the basic elements of the theory and the observations are proposed. 
   }   
   {}

   
   \titlerunning{Concurrent X-ray, NIR, Sub-mm, and GeV $\gamma$-ray observations of \sgra}
\authorrunning{G. Trap et al.}
   \maketitle
 

\section{Introduction}

The tracing of stellar orbits in the heart of the Milky Way since the early 90's has definitely established the presence of a central high concentration of mass, $\sim$$4\times10^6$~$M_{\odot}$, confined to a region no bigger than the Solar System \citep{ghez08,gillessen09}. This is the best evidence to date for the existence of a supermassive black hole at the Galactic center \citep[GC, see][for a comprehensive review]{melia07}. 
The relative proximity of this black hole {\citep[8~kpc,][]{reid09} } has enabled thorough observations of its weak point-like radiative manisfestation, Sagittarius~A$^{\star}$ (\sgra), all accross the electromagnetic spectrum. In spite of a high interstellar extinction, wich precludes detections at optical and ultraviolet wavelengths, \sgra\ has now been positively detected in radio \citep[from cm to sub-mm wavelengths,][]{balick74,zylka88}, near-infrared \citep[NIR,][]{genzel03}, and X-rays \citep{baganoff03a}. 
In the vast $\gamma$-ray domain, point-like sources coincident with the position of \sgra\ have been found at the 10~keV \citep[\sgraigr,][]{belanger04}, GeV \citep[\sgrafermi,][]{abdo10}, and TeV  \citep[\sgrahess,][]{aharonian04} levels, though their association with the supermassive black hole is less evident.

One striking feature of \sgra\ is its temporal variability. Albeit much more quiet than typical active galactic nuclei (AGNs), \sgra~displays rapid flares from time to time, in X-rays and NIR, thus temporarily enhancing its luminosity by factors up to $\sim$160 on timescales of 1--2~hr. X-ray flares have been discovered at a rate of about one event per day by the \chandra~\citep{baganoff01} and \xmm~\citep{goldwurm03} satellites, while NIR flares seem to recur more frequently (2--6 events per day) as observed by the \vlt~\citep{genzel03}, the \keck~\citep{ghez04}, the \hst~\citep{yusef-zadeh06a}, and \subaru~\citep{nishiyama09}.

These flares usually present short timescale variations pointing to an emitting region with a size on the order of 10\,$R_{\rm S}$ or less, where $R_{\rm S}$ is the Schwarzschild radius of the black hole. They have thus offered new probes into the central engine and received considerable attention via extensive multi-wavelength campaigns since their discoveries. On the one hand, such campaigns revealed that all X-ray flares seem to have a simultaneous NIR counterpart \citep{eckart04,eckart06a,eckart08a,hornstein07,yusef-zadeh06a,yusef-zadeh09, dodds09}, while not all NIR flares have detectable X-ray counterparts \citep{hornstein07}. Considering also the linear polarization properties of NIR flares \citep[e.g.,][]{eckart06b}, a general consensus emerged in which the origin of NIR flares was ascribed to synchrotron radiation of transiently accelerated electrons and X-ray flares to the corresponding inverse Compton emission, especially synchrotron self-Compton (SSC) \citep[e.g.,][]{liu06a,liu06b}. However, the recent study of an exceptionally bright X-ray/NIR flare detected on April $4^{\rm th}$, 2007, challenged the SSC model \citep{dodds09,trap10}. This proved the necessity to measure contemporaneous spectra (with good statistics) in infrared and in X-rays, to constrain the overall spectral energy distribution (SED) of the flares. Yet, past attempts to obtain NIR spectral information on \sgra~have given various and sometimes conflicting results \citep[e.g.,][]{gillessen06,hornstein07}. 

On the other hand, broadband campaigns found indications that contemporaneous X-ray/NIR flares could be accompanied by delayed radio activity \citep{yusef-zadeh06a,eckart06a}. This was interpreted as a sign of expanding plasmoids in the vicinity of the supermassive black hole, releasing optically thin and thick synchrotron emission whose peak evolves with frequency \citep{vanderlaan66}. 
Recently, \citet{marrone08} and \citet{yusef-zadeh08} clearly detected an increase in the mm and sub-mm emission, using the \textit{Sub-Millimeter Array} (\textit{SMA}) and the \textit{Caltech Sub-millimeter Observatory} (\textit{CSO}) respectively, following an X-ray/NIR flare detected by \chandra~and the \keck\ observatories.  
In the frame of an SSC/expanding blob model, \citet{marrone08} argued that  there should be an observable correlation between the X-ray to NIR peak ratio and the sub-mm time delay, which could help test this picture.

{We present here a new X-ray, NIR, and sub-mm campaign involving the \textit{X-ray Multi-Miror (XMM)-Newton} satellite, the \textit{Very Large Telescope (VLT)}, and the \textit{Atacama Pathfinder EXperiment (APEX)} observatory, respectively. The data obtained enable us to investigate the aforementioned issues by measuring \sgra's time variability and spectrum across a broad range of wavelengths. 
Concurrent public GeV $\gamma$-ray data from the \fermi\ satellite are also examined.}
Additional radio data will be reported elsewhere.

This paper is organized as follows. 
In Sect.~\ref{obs_section}, we give brief summaries of the data acquisition and reduction of each telescope. In Sect.~\ref{light curves}, we describe the respective light curves of \sgra, identify new flares, and compare their properties to past detections. Sect.~\ref{color_section} is dedicated to the NIR color measurements performed on one particular flare, while Sect.~\ref{discussion} discusses possible phenomenological interpretations of the light curves and SEDs of all the newly detected flares.

 \begin{figure*} 
   \centering
\includegraphics[trim=0cm 7.5cm 0cm 6.5cm, clip=true, width=18.5cm]{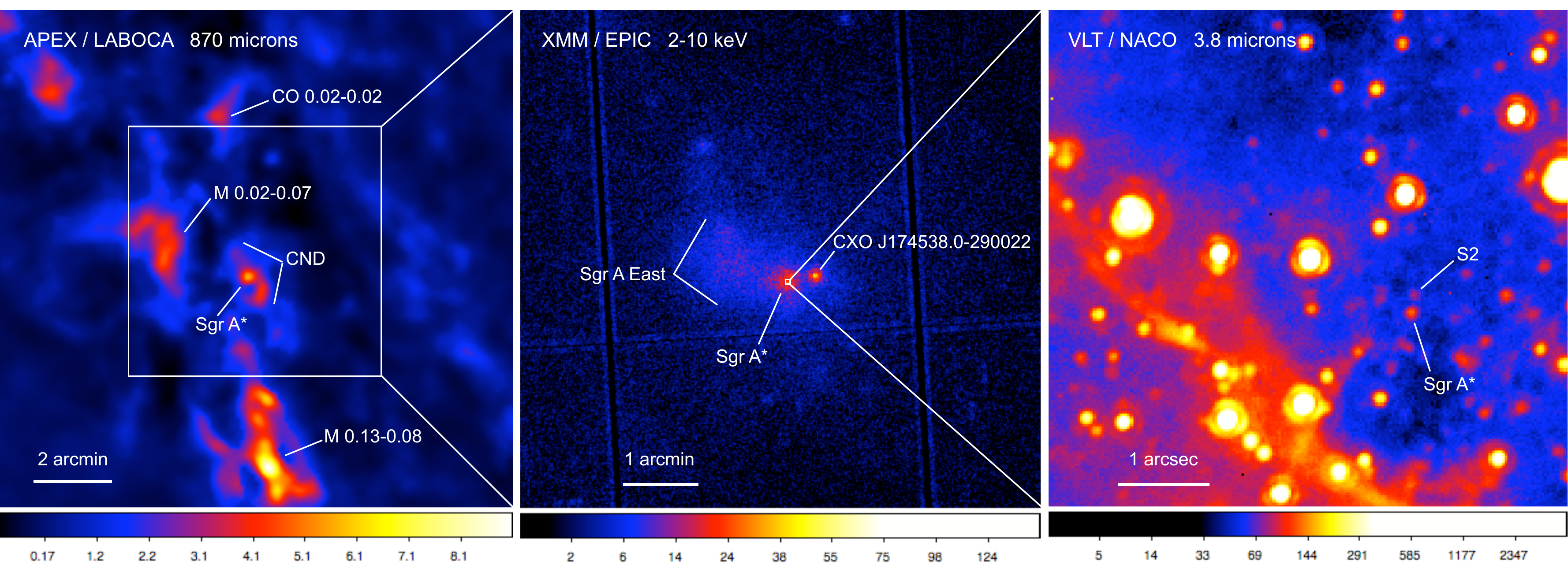}
   \caption{Multi-wavelength images of the Galactic center in early April 2009. North is to the top, East to the left, and the Galactic plane runs from upper left to bottom right. The sub-mm and X-ray images (\textit{left and middle panels}) are summed over one or several observations, whereas the $L'$-band NIR image (\textit{right panel}) corresponds to the summit of Flare~A.
   Color scales are given in units of Jy\,beam$^{-1}$, counts\,pixel$^{-1}$, and ADU, from left to right, respectively.}
   \label{ima}
\end{figure*}


\section{Observations and data analysis}
\label{obs_section}
We observed the GC in early April 2009 in the context of a large multi-wavelength campaign that was initiated by an \xmm/\vlt\ joint program and followed by an \apex\ monitoring scheduled accordingly. 
A detailed observation journal for this set of instruments is given in Tab.~\ref{log}.
\fermi\ data too were recorded at the epoch of this campaign as part of the ongoing monitoring of the GeV $\gamma$-ray sky. 
We present these data and their respective treatments below.

\begin{table}
\begin{minipage}[t]{\columnwidth}
\caption{Observation log.}
\label{log}
\centering
\renewcommand{\footnoterule}{}  
\begin{tiny}
\begin{tabular}{lcccc}
\hline \hline

\textsc{Orbit}		&	\textsc{ObsID}       & 	\textsc{Start date}    & \textsc{Exp.} & \textsc{Band} \\
~ & ~ & [UTC] &  [ks]  & ~ \\
\hline
\\
\smallskip  
 
                   &                          & \xmm/PN                                                               &              &                   \\
\hline
1705		& 554750401	& 2009 Apr 1, 01:18:30				& 37.8  	& 2--10~keV \\
1706		& 554750501	& 2009 Apr 3, 01:56:31				& 42.0  	& 2--10~keV \\
1707		& 554750601	& 2009 Apr 5, 03:53:34				& 32.4  	& 2--10~keV \\

\hline
\\
\smallskip  
                   &                          & \vlt/NACO                                                               &              &                   \\
\hline
\dotfill	& 082.B-0952A		& 2009 Apr 1, 06:46:59				& 11.5  	& 3.8~$\mu$m ($L'$) \\
\dotfill	& 082.B-0952A		& 2009 Apr 3, 07:25:41				& 9.7  	& 2.1~$\mu$m ($K_{\rm s}$)  \\
\dotfill	& 082.B-0952A		& 2009 Apr 3, 07:28:42				& 9.7  	& 3.8~$\mu$m ($L'$) \\
\dotfill	& 082.B-0952A		& 2009 Apr 5, 08:44:10				& 6.6  	& 3.8~$\mu$m ($L'$) \\

\hline
\\
\smallskip  
                   &                          & \apex/LABOCA                                                               &              &                   \\
\hline
\dotfill	& 183.B-0100A		& 2009 Apr 1, 05:31:43				& 30.7  	& 870~$\mu$m \\
\dotfill	& 183.B-0100A		& 2009 Apr 3, 11:07:43				& 8.2  	& 870~$\mu$m \\
\dotfill	& 183.B-0100A		& 2009 Apr 5, 06:40:10				& 25.6  	& 870~$\mu$m \\

\hline
\end{tabular}
\end{tiny}
\end{minipage}
\end{table}

\subsection{XMM-Newton (2--10 keV)}
\label{xmm}

The X-ray satellite \xmm\ \citep{jansen01} pointed toward the GC on three occasions between April 1\st\ and Apr 5\ith\ 2009 ($\sim$$3\times40$~ks). In the following, we will consider the data recorded with the \texttt{Medium} filter by the \textit{European Photon Imaging Camera (EPIC)}, i.e. the PN \citep{struder01} and the MOS\,1--2 cameras \citep{turner01} instruments, all operated in \texttt{Prime Full Window} mode. The data were reduced and analyzed with the \textit{Science Analysis Software (SAS)} package (version 8.0.0)  and the latest calibration files available at the time of the analysis. They were not contaminated by soft proton flares from solar activity. Starting with the original Observation Data Files, event lists were generated through the \texttt{emchain} and \texttt{epchain} routines, and filtered
with \texttt{PATTERN\,<=\,4} or \texttt{12} (for PN and MOS respectively), to ensure the selection of X-rays
and rejection of cosmic rays. In order to prevent the use of X-ray photons that fell
on bad pixels, we also applied the \texttt{\#XMMEA\_EP} and \texttt{\#XMMEA\_EM} filters on the event lists.
Finally, artefacts from the calibrated, and concatenated datasets and events near CCD
gaps were discarded by setting \texttt{FLAG\,==\,0} as selection criterion. 
We note that the high interstellar absorption toward the GC ($N_{\rm H}\approx10^{23}$~\cm) prevents the use of photons below $\sim$2 keV, so that the following count rates, fluxes, and luminosities will refer to the 2--10~keV~band.

The image obtained by the PN detector and reproduced in Fig.~\ref{ima} (middle panel), reveals the usual constituents of the Galactic nucleus visible via \xmm: extended thermal emission from the shell of the supernova remnant Sgr\,A\,East \citep{sakano04} and point-like emission conincident with the position of \sgra. In addition, these observations allowed to locate another point source at only $\sim$27$''$ from \sgra, which we identify with the weak transient CXOGC\,J174538.0--290022 \citep[$\sim$$2\times10^{34}$ \ergpersec\ at 8~kpc,][]{ponti09}.

To search for time variability of \sgra, we first fitted the positions of the point sources by running the \texttt{edetect\_chain} procedure. A light curve of \sgra\ was then extracted from a $10''$ radius circular region centered on the fitted position in each EPIC detectors. These light curves were background subtracted, with the background regions selected following the prescriptions given in \cite{kirsch05} and the counts scaled to the same extraction area as \sgra.
Finally, the 2--10~keV background subtracted PN, MOS\,1, and MOS\,2 light curves were summed and rebinned with a sampling of 600~s for the 1\st\ and 3\rd\ observations, and 400~s for the 2\nd.
Note that, as resolved by previous \chandra\ deep observations, the persistent X-ray emission from a $10''$ radius region centered on \sgra, not only contains the X-ray counterpart of \sgra, but also includes persistent compact sources such as the star group IRS\,13 \citep{baganoff03a} and the pulsar wind nebula candidate \pwin\ \citep{wang06}, possible transient point sources \citep{muno05a}, as well as a diffuse component \citep{muno03a}.

\subsection{VLT (2.1 \micron\ and 3.8 \micron)}

The three X-ray observations described in the last paragraph were partly covered by NIR observations with the NACO instrument, {NAOS \cite[Nasmyth Adaptive Optics System,][]{lenzen03} + CONICA \cite[COud\'e Near Infrared CAmera,][]{rousset03}}, mounted on the fourth unit telescope, Yepun, at ESO's \textit{VLT} in Paranal, Chile.
These observations were divided into three half nights (April 1\st, 3\rd, and 5\ith\ 2009), which all started in the middle of the night and were extended as long as possible until sunrise. 

We acquired data solely in the $L'$-band (3.8~\micron) during the 1\st\ and 3\rd\ nights, while during the 2\nd\ night, atmospheric conditions were favorable to also collect images in the $K$-short ($K_{\rm s}$) band at 2.1~\micron. During this latter night, we attempted a novel filter switching technique with NACO in order to study the color of \sgra\ from broadband quasi-simultaneous NIR measurements. Up to now, it has only been possible to investigate the spectrum of \sgra\ at the \vlt\ through the integral field spectrograph SINFONI over relatively narrow bands: the $H$+$K$-band \citep[$\sim$1.7--2.4~\micron,][]{eisenhauer05} and the $K$-band \citep[$\sim$2.0--2.4~\micron,][]{gillessen06}. 
Using NACO, we obtained for the first time spectral data over a wider wavelength range and with better Strehl ratios, by constantly cycling back and forth between $K_{\rm s}$ and $L'$ photometric filters throughout the night. We carried out successive full switches (switching $K_{\rm s}$-  to $L'$-band, acquiring two $L'$-band images, switching $L'$- to $K_{\rm s}$-band and acquiring two $K_{\rm s}$-band images) on a timescale of only $\sim$5--5.5~min. All the data examined here were collected with the supergiant IRS\,7 as natural guide star and the infrared wavefront sensor to close the adaptive optics loop of NAOS.
Images were individually sky-subtracted, flat-fielded, and corrected for dead/hot pixels (see Fig.~\ref{ima}, right panel). 

In Fig.~\ref{n1}, \ref{n2}, and \ref{n3} we display the light curves of \sgra\ corrected for extinction with the usual values $A_{K_{\rm s}}=2.8$~mag and $A_{L'}=1.8$~mag, in order to allow comparison with previous works (e.g., \citealt{genzel03,dodds09}). They were obtained via aperture photometry and calibrated
with the fluxes of nearby stars of known and stable brightness: all fluxes were scaled so that the median flux of the relatively isolated star S\,65 always corresponds to an apparent $K_{\rm s}$ magnitude of 13.7~mag \citep{gillessen09} and the median flux of IRS\,16\,NW to an apparent $L'$ magnitude of 8.38~mag \citep{viehmann05}.
More details on the 2\nd\ night light curves are given in Sect.~\ref{color_section}. The duration of the 3\rd\ night light curve is reduced compared to the other two due to a technical problem with the telescope. We stress that, at the time of these observations, \sgra\ was confused with the star S\,17, whose contribution to the light curves is expected to be a constant value of $\sim$5.4~mJy and $\sim$3.5~mJy in the $K_{\rm s}$-  and $L'$-band, respectively \citep[see S\,17 magnitudes reported in][corrected with the aforementioned values of $A_{\lambda}$]{ghez05b}.

\subsection{APEX (870 \micron)}

In parallel to the X-ray and NIR surveys, a total of $\sim$20~hr of sub-mm GC data were taken at 870~\micron\ (345~GHz). We used the Large \textit{APEX} bolometer camera \citep[LABOCA,][]{siringo09},  on the \textit{APEX} telescope \citep{gusten06}
at Llano Chajnantor, Chile. LABOCA is an array of 295 composite 
bolometers operating in the 345~GHz transmission window. The angular resolution (FWHM) of each beam is $\sim$$19''$ for a bandwidth of $\sim$60~GHz.
LABOCA's layout leads to a double beam spaced distribution of the individual 
beams in a hexagonal configuration over a $11.4'$ field of view (FOV).

The observations performed on the 1\st\ and the 5\ith\ of April spanned $\sim$8.5 and $\sim$7.5~hr, respectively, with an interruption of 1~hr and 15~min between UT 8:45 and 10:00. During this time interval, the GC passed above the telescope's elevation limit of $\sim$80$^\circ$.
Observations on April 3\rd\ only covered the period after this GC transit 
(UT 11:00 -- 13:30). Note that the \apex\ and \vlt\ sites have basically the same longitude, but, contrary to the \vlt, \apex\ could continue observing past sunrise, till the elevation of the GC was too low to allow for reliable monitoring purposes (elevation between
30$^\circ$ and 40$^\circ$, depending on weather conditions). The precipitable water vapor during the three runs was $\sim$2, $\sim$0.9, and 0.2--0.4~mm, respectively. 

The GC was observed using ``on the fly'' maps with scanning
angles of $-$15$^\circ$, 0$^\circ$, or $+$15$^\circ$ relative to the orthogonal
direction across the Galactic plane. The maps were built with a scanning
speed of 2~arcmin\,s$^{-1}$ and steps orthogonal to the scanning direction of $30''$.
The total integration time of each map was 315~s. Each GC 
map was followed by observations of G\,10.62 and G\,5.89, standard LABOCA secondary calibrators. The atmospheric zenith opacity was determined via skydips every 1--2~hr. 
With this observing setup, \sgra\ flux was monitored every $\sim$8~min.

The data were reduced using the BoA software package. Reduction steps on the 
bolometer time series include temperature drift correction based on two 
``blind'' bolometers, 
flat-fielding, calibration, opacity correction, correlated noise removal on 
the full array as well as on groups of bolometers related by the wiring and 
in the electronics, flagging of bad bolometers, and finally de-spiking. Each reduced 
scan was then gridded into a spatial intensity and weighting map. These
reduction steps were applied to the GC and both calibrators
scans. Flux calibration was achieved using G\,10.62, which has a known flux density
of $33.4\pm2.0$~Jy for LABOCA \citep{siringo09}. The absolute 
calibration accuracy is roughly 10\%. To infer the relative calibration
error, we applied the calibration curve of G\,10.62 to G\,5.89 and determined
the dispersion of the measured G\,5.89 fluxes over the observing period. 
This yielded a relative calibration accuracy of $\sim$3\%.

For the determination of \sgra's light curve we first generated
a model of the sub-mm emission of the region by co-adding all 
calibrated and pointing drift corrected GC maps. In this high signal to 
noise map (see Fig.~\ref{ima}, left panel), we note the presence of several molecular clouds in accordance with previous sub-mm GC surveys \citep{pierce00}, and the point source \sgra, which was fitted with a gaussian and later subtracted from the co-added map. We finally re-reduced all 
GC scans and subtracted the model signal from the time serie of 
each bolometer. The resulting maps only contain the point source \sgra.
The light curves were eventually constructed by 
fitting a gaussian to each individual map (see Fig.~\ref{n1}, \ref{n2}, and \ref{n3}). The error bars were estimated through the RMS of the calibrator G\,5.89 ($\sim$2.7~\%). Note that \sgra\ does not suffer from any significant interstellar extinction at 870~\micron. 

\subsection{Fermi (0.1--200 GeV)}

{The \fermi\ space observatory carries the Large Area Telescope \citep[LAT,][]{atwood09}, a high-energy $\gamma$-ray pair-conversion telescope covering the $\sim$30 MeV to $\sim$300~GeV band with a wide FOV (2.4~sr). It operates almost as an all-sky monitor and can provide regular $\sim$50~min uninterrupted exposures on the GC.
Using the public \fermi\ data archive, we explored the periods covered by the aforementioned observations of April 2009, and searched for a potential $\gamma$-ray flaring activity correlated with variability in other spectral domains.}

The LAT data were analyzed with the \fermi\ Science Tools  and following the analysis threads distributed by the \fermi\ Science Support Center. We created events cube with the \texttt{gtselect} routine by selecting only $\gamma$-ray photons (``diffuse'' class events, N\degree3). We then used \texttt{gtmktime} and the ``spacecraft'' file which keeps track of the satellite's attitude as a function of time, to perform good time intervals selections. A zenith-angle cut was thereby applied ($z_{\rm max}=105$\degree), rejecting $\gamma$-rays emanating from Earth albedo. The light curves were derived via aperture photometry (\texttt{gtbin}) with a circular extraction region of radius 1\degree, centered on the position of \sgra. The fluxes thus measured include the diffuse $\gamma$-ray emission associated with cosmic ray interactions in GC molecular clouds.
The number of counts in each bin, $n$, being governed by Poissonian statistics, 1$\sigma$ (84\% confidence level) error bars were approximated by $1+\sqrt{n+0.75}$ \citep{gehrels86}.
Finally, we computed the exposures of each time bin in s$^{-1}$\,cm$^{-2}$ with \texttt{gtexposure}. The resulting 0.1--200~GeV light curves are presented in the top panels of Fig.~\ref{n1}, \ref{n2}, and \ref{n3}, with a time bin of 600~s (10~min). 

\begin{figure} 

   \centering
   \includegraphics[trim=0.1cm 0cm 0.82cm 0cm, clip=true, width=9cm]{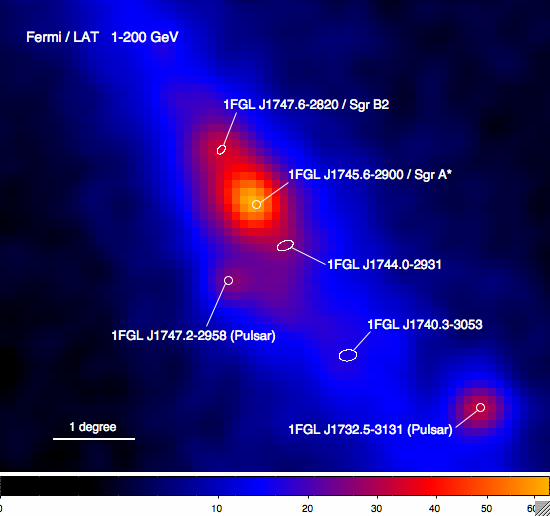}
   \caption{Map of the Galactic center built with 1--200~GeV $\gamma$-ray photons collected by the LAT instrument during the 1\st\ year of the \fermi\ mission (Aug 2008/Aug 2009). North is to the top, East to the left, and the Galactic plane runs from upper left to bottom right. Each pixel has a size of 0.1\degree\ and the image was smoothed with a gaussian kernel of 0.15\degree\ radius. The color bar is expressed in counts and the positions of the point sources were taken from \cite{abdo10}.}
   \label{ima_fermi}
\end{figure}

\begin{figure} 

   \centering
      \includegraphics[trim=1.73cm 4.5cm 1cm 3.cm, clip=true, width=9.5cm]{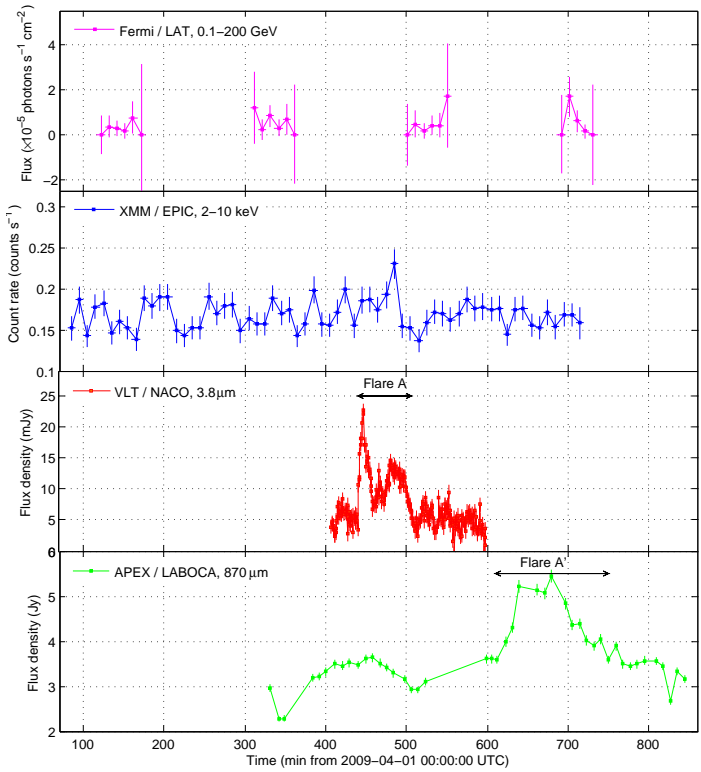}

   \caption{Light curves of \sgra\ and its associated backgrounds (see text for details) during the 1\st\ observation.}
   \label{n1}
\end{figure}

\begin{figure} 

   \centering
      \includegraphics[trim=1.73cm 4.5cm 1cm 3.cm, clip=true, width=9.5cm]{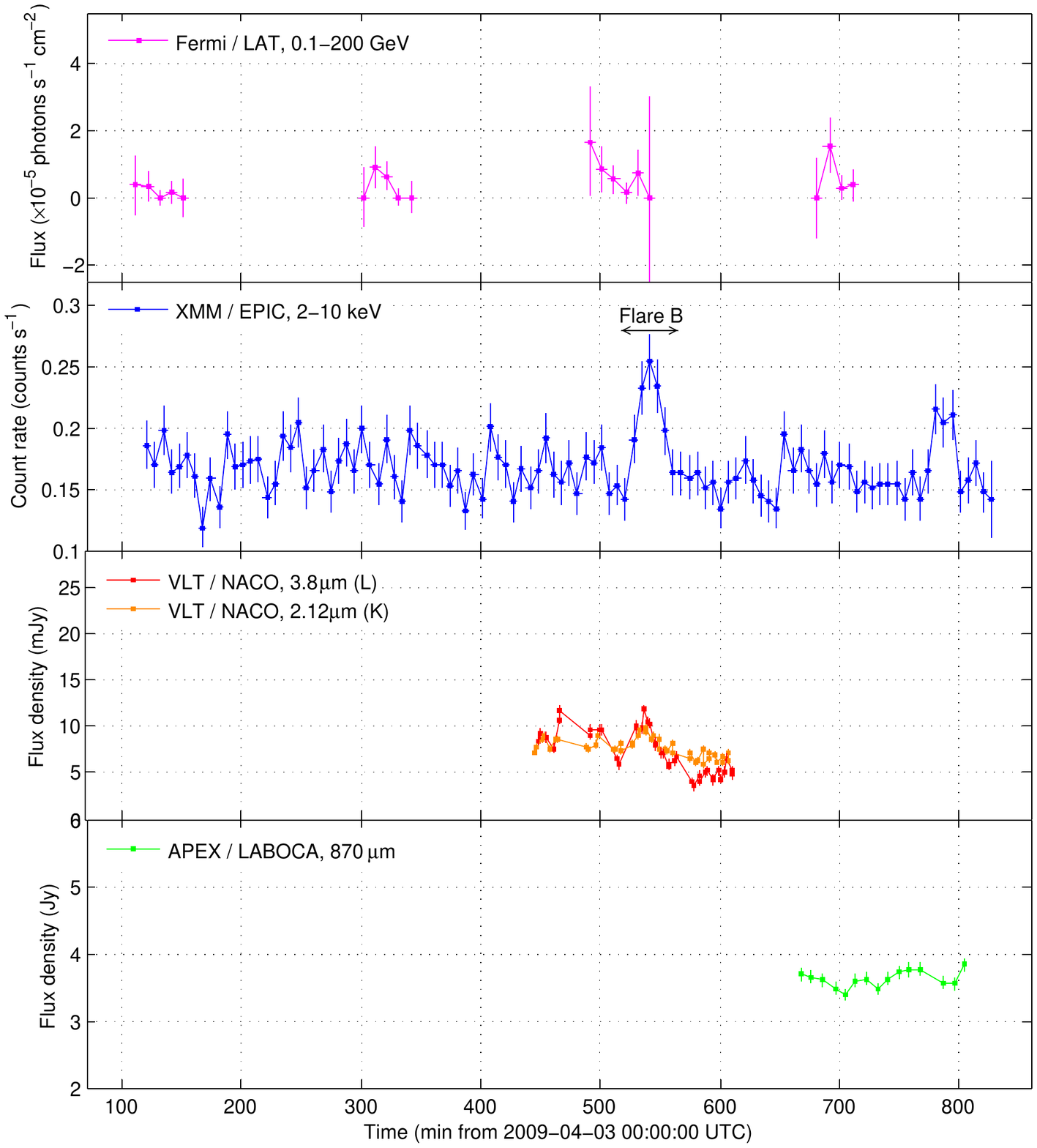}
   \caption{Same as Fig.~\ref{n1} for the 2\nd\ observation.}
   \label{n2}
\end{figure}

\begin{figure} 

   \centering
      \includegraphics[trim=1.73cm 4.5cm 1cm 3.cm, clip=true, width=9.5cm]{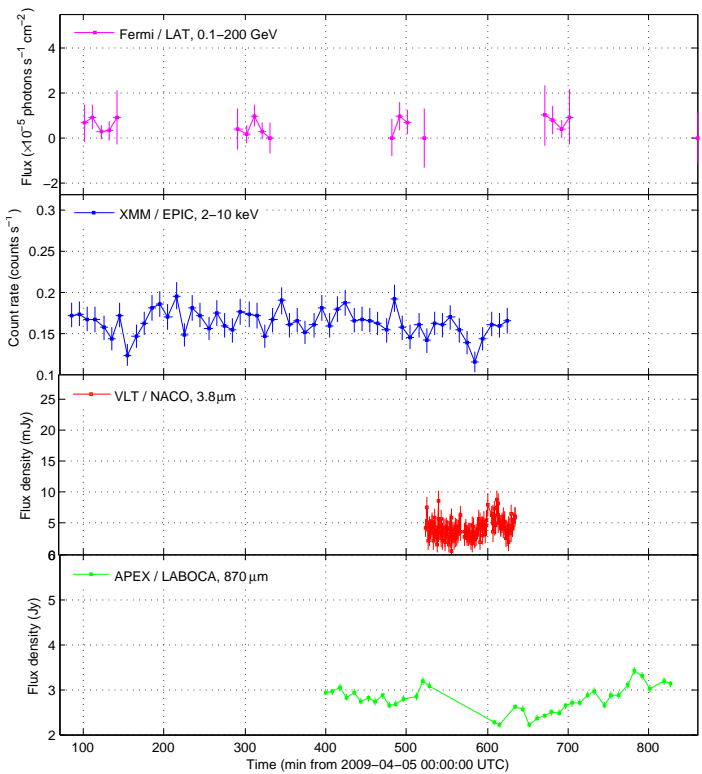}
   \caption{Same as Fig.~\ref{n1} for the 3\rd\ observation.}
   \label{n3}
\end{figure}

For illustration purposes, we produced an average LAT image of the GC (Fig.~\ref{ima_fermi}) using all the data from the 1\st\ year of full sky survey, from August 2008 to August 2009. Only high energy photons from the 1--200~GeV band were considered to optimize the single photon angular resolution ($<0.6$\degree\ at energies $>1$~GeV).

\section{Light curves and flares}
\label{light curves}
In this Section, we discuss the properties of the multi-wavelength light curves and compare them with the relevant literature.

\subsection{NIR}
\label{nir_para}

The 1\st\ night of observations was marked by a strong NIR flare detected in the $L'$-band between 440--515~min past midnight (Fig.~\ref{n1}), hereafter referred to as ``Flare~A''. 
It was brighter than the star S\,2 (see the pale red dot north of \sgra\ in Fig.~\ref{ima}, right panel) and peaked at $19\pm1$~mJy (background subtracted), which makes it the 2\nd\ brightest $L'$ flare covered in X-rays and one of the few powerful NIR flares so far detected that are on the order of 20~mJy (in all NIR bands considered) or beyond \citep{dodds09,eckart08a,eckart08b, meyer06a,meyer07}.
It is worth noting that Flare~A has a complex non-symmetrical morphology, characterized by a very sharp rise (a continuous gain of $\sim$19~mJy in $\sim$7~min) followed by a slower and {non-monotonic} decay. This rise time is one of the fastest ever reported and is actually rather unusual among strong NIR flares. It hence places a stringent upper limit on the size of the \textit{overall} flare emitting region of $\sim$10\,$R_{\rm S}$ (or $\sim$1~AU), with a causality argument ($R\leq c\,\Delta t$) and considering a mass of $4\times10^6$~$M_\odot$ for the black hole.

During the 2\nd\ night, the time resolution of the light curves was not as good as during the other nights, given the continuous cycling through the $K$ and $L$ filters, and hence makes the identification of flares more difficult. \sgra\ seems to have been continuously variable in both bands {at a significant level (see Fig.~\ref{color_katie}, top panel, and \textsection~\ref{color_section})}, at least during the first two thirds of the observation. 
Our X-ray coverage allows us to clearly isolate a new eruption---Flare~B---between $\sim$520 and $\sim$560~min after 2009--04--03 00:00:00 UTC though (see paragraph~\ref{paragraph_xrays} and Sect.~\ref{color_section}). This time, the shape of the flare's light curve is more symmetrical, with no time delay between the peaks of the $K_{\rm s}$- and $L'$-bands larger than $\sim$5~min, the time resolution of the light curve. The peak fluxes of Flare~B are $3.7\pm0.5$ and $8.2\pm0.5$~mJy (background subtracted) in the $K_{\rm s}$- and $L'$-band, respectively. 

No significant flare was recorded during the 3\rd\ night, which had an observed mean flux density of $\sim$4~mJy.

\begin{table*}

\renewcommand{\footnoterule}{}  
\begin{minipage}[t]{18cm}

\caption{Properties of the X-ray/NIR flares observed so far.}
\centering
\begin{tiny}
\label{tab_nirx}

\begin{tabular}{lrrlcccl}
\hline \hline
\vspace{0.05cm}

\textsc{References} \footnote{References: 
[1]~This work, 
[2]~\cite{porquet08},
[3]~\cite{dodds09}, 
[4]~\cite{yusef-zadeh09}, 
[5]~\cite{belanger05},
[6]~\cite{yusef-zadeh06a}, 
[7]~\cite{hornstein07}, 
[8]~\cite{eckart08a},  
[9]~\cite{eckart06a}, 
[10]~\cite{eckart04}. }
&
\textsc{Date}       
& \textsc{Instruments}
& \textsc{NIR filter} \footnote{The \hst\ 1.6 and 1.7 \micron\ pass-bands correspond to the NICMOS filters F160W and F170M, respectively.}
& $L_{\rm X}/L_{\rm quies}$ \footnote{Peak X-ray luminosity (2--10~keV) in units of the quiescent level \citep[$\sim$$2 \times 10^{33}$~\ergpersec,][]{baganoff03a}. `$>$' and `$<$' signs were added when the peak flux was missed or not simultaneous to the NIR peak.}  
& $F_{\rm NIR}$ \footnote{Background subtracted peak NIR fluxes in the bands indicated in the fourth column, dereddened assuming  $A_{\rm 1.6\,\mu m}=4.5$~mag, $A_{\rm 1.7\,\mu m}=5.03$~mag, $A_{\rm 2.1\,\mu m}=2.8$~mag, and $A_{\rm 3.8\,\mu m}=1.8$~mag.} 
& $F^{L'}_{\rm NIR}$ \footnote{Equivalent NIR flux in the $L'$-band extrapolated assuming a power-law spectral index $\alpha=-1$ (see Sect.~\ref{color_section}).}
& \textsc{Id.}   
   \\

~ & ~ & ~ & ~ & ~ & [mJy]  & [mJy] & ~\\

\hline
[1] \dotfill & 1 Apr 2009								& \xmmshort~/~\vlt & $3.8$ \micron~($L'$)	& $<7$	& $19$ 	&  \dots & A \\

[1] \dotfill & 3 Apr 2009 								& \xmmshort~/~\vlt & $3.8$ \micron~($L'$)	& $10$	& $8$ 	&  \dots & B \\

[2,3] \dotfill& 4 Apr 2007 				& \xmmshort~/~\vlt & $3.8$ \micron~($L'$)	& $103$	& $26$ 	& \dots & C  \\

[2,4] \dotfill& 4 Apr 2007 		& \xmmshort~/~\hst & $1.7$ \micron			& $26$	& $7$ 	& $16$ & D  \\

[2,4] \dotfill& 4 Apr 2007 		& \xmmshort~/~\hst & $1.7$ \micron			& $37$	& $4$ 	& $9$ & E  \\

[2,4] \dotfill& 2 Apr 2007 		& \xmmshort~/~\hst & $1.7$ \micron			& $14$	& $10$ 	& $22$ & F \\

[5,6] \dotfill & 31 Aug 2004	 	& \xmmshort~/~\hst & $1.6$ \micron			& $<40$	& $10$ 	& $24$ & G \\

[7] \dotfill & 31 Jul 2005	 					& \chandra~/~\keck & $3.8$ \micron~($L'$)	& $<1.2$	& $15$ 	& \dots & H \\

[7] \dotfill & 17 Jul 2006	 					& \chandra~/~\keck & $3.8$ \micron~($L'$)	& $20$	& $>5$ 	& \dots & I  \\

[8] \dotfill  & 30 Jun 2005	 					& \chandra~/~\vlt & $2.1$ \micron~($K_{\rm s}$)		& $3$	& $14$ & $4$ & J \\

[9] \dotfill & 7 Jul 2005	 						& \chandra~/~\vlt & $2.1$ \micron~($K_{\rm s}$)		& $15$	& $11$ & $3$ & K  \\

[10] \dotfill  & 19 Jun 2003	 						& \chandra~/~\vlt & $2.1$ \micron~($K_{\rm s}$)		& $3$	& $>7$ & $>2$ & L  \\

\hline
\end{tabular}
\end{tiny}
\end{minipage}
\end{table*}

\subsection{X-rays}
\label{paragraph_xrays}

No {prominent X-ray flare ($>10$ times the quiescent level)} was detected in April 2009. We concentrate, hereafter, on the significant peaks ($>$3$\sigma$) of the X-ray light curves detected in parallel to the NIR flares mentioned above.

{ Flare~A coincided with a very little increase of the \xmm/EPIC cameras count rate (see Fig.~\ref{n1}, top and middle panels). Though barely significant (3.5$\sigma$), this could be the X-ray counterpart to Flare~A and thus would be the faintest X-ray flare so far identified with \xmm, all previous ones being, at least, a factor 2 brighter \citep{goldwurm03,porquet03, belanger05,porquet08}. 
 When considering the overall flare, the X-ray excess occurs in the middle of the NIR flare, which is consistent with the flares that have been entirely covered in NIR and X-rays over the past \citep{eckart06a,eckart08a,dodds09}.
In view of the low statistics of this X-ray signal, we cannot extract proper spectral information. }Assuming a typical photon spectral index, $\Gamma=2$ (with $\mathrm{d}N/\mathrm{d}E\propto E^{-\Gamma}$~\phperseccmkev), and a typical column density, $N_{\rm H}=10^{23}$~\cm, we estimate through WebPIMMS that Flare~A reached an unabsorbed, background subtracted\footnote{An average level of quiescent count rate including the constant contibution of neighboring sources (\textsection~\ref{xmm}), was removed as in \cite{goldwurm03}.}, flux  of $2.2\times10^{-12}$~\ergperseccm\ in the 2--10~keV band. This translates to a peak luminosity of $1.7 \times 10^{34}$~\ergpersec\ at 8~kpc, i.e. $\sim$7 times the quiscent level \citep[$2.3 \times 10^{33}$~\ergpersec,][]{baganoff03a}.
This very low significance X-ray flare associated with a relatively bright NIR flare is quite surprising in regard to the bright X-ray flare \citep[$\sim$100 $\times$ the quiescence,][]{porquet08} detected simultaneously to a strong $L'$ NIR flare ($\sim$30~mJy) in April 2007 \citep{dodds09}. It is however somewhat reminescent of the X-ray non-detection by the \chandra\ observatory of a $\sim$15~mJy $L'$ flare observed in July 2005 by \citet{hornstein07}. Yet, we stress that the point spread functions (PSFs) and spectral responses of the \chandra/ACIS and \xmm/EPIC instruments differ, which prevents straightforward comparisons. Actually, the fact that the EPIC's PSF {is broader than ACIS's} makes the \xmm\ satellite less sensitive than \chandra\ to soft ($\Gamma \approx 2$) and faint X-ray flares. The background subtracted peak \xmm\ count rate of Flare~A converts, indeed, to a \chandra/ACIS-I rate of $\sim$0.034~\ctspersec\ in the 2--8~keV band. This would have lead to a clear \chandra\ detection, even more significant than two other faint X-ray/NIR flares observed with \chandra\ by \citet{eckart04,eckart08a}.

Flare~B, on the other hand, was more significantly detected by \xmm\ and observed to last longer ($\sim$40~min), in sync with the NIR light curve (see Fig.~\ref{n2}, top and middle panels).
Using the same method as above, we estimate its background subtracted and unabsorbed peak flux is $3.2\times10^{-12}$~\ergperseccm\ in the 2--10~keV band, i.e. $2.4 \times 10^{34}$~\ergpersec\  ($\sim$10 $\times$ the quiescence). Note that, with \chandra, it would have been detected at a rate of 0.049~\ctspersec\ (2--8~keV), which is still faint compared to the other X-ray flares with a NIR counterpart known as we discuss below. 
{
The fact that the NIR flux was relatively high about 100 min before Flare~B with no simultaneous detectable X-ray emission is striking and could be another example of a ``NIR only flare'' \citep{hornstein07} or indicate that NIR flares can start earlier than the X-ray ones \citep{dodds09}. 
}

Comparing Flare~A and B, it is manifest that the ratio of the peak NIR flux over the peak X-ray flux is highly variable from one flare to another. It is known that the NIR flares occurence rate is higher than the X-ray one, so it was expected that there should exist NIR flares with very little or even no
detectable X-ray emission. But a somewhat counterintuitive fact that emerges now is that
the brightest NIR flares are not necessarily accompanied by bright X-ray flares.
Indications of that result had been obtained with the analysis of the July 2005 and July 2006 flares in \citet{hornstein07}, although the peak NIR flux of the July 2005 event was missed. Similarly, \citet{yusef-zadeh09} showed in their Fig.~21 (a) and (b), with a consistent \xmm/\hst\ dataset, that the NIR to X-ray ratio fluctuates.

\begin{figure} 

   \centering
    \includegraphics[trim=2.cm 6.5cm 3cm 6.5cm, clip=true, width=9cm]{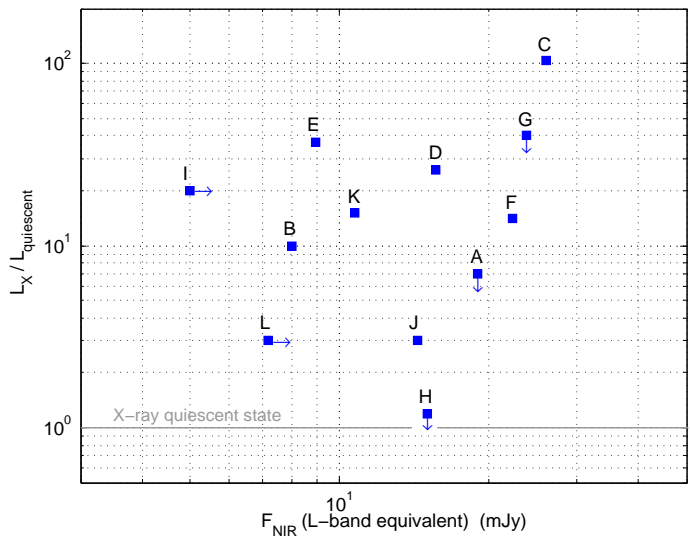}
   \caption{Relation between the X-ray and NIR peak fluxes of all the flares so far covered with multi-wavelength observations. See Tab.~\ref{tab_nirx} for details about each individual flare. Events A and B are the ones reported in this work. X-ray upper limits and NIR lower limits are indicated by down pointing and right pointing arrows, respectively.}
   \label{fig_nirx}
\end{figure}

To allow better comparisons between flares concurrently observed in X-rays and NIR, and visualize their disparities, we give an inventory of all the results published thus far in Tab.~\ref{tab_nirx} and plot the X-ray peak fluxes Vs. peak NIR fluxes in Fig.~\ref{fig_nirx}.  
Despite a relatively easy comparison between the measurements conducted by \xmm\ and \chandra, NIR comparisons between various authors are more difficult because of different observational setups and flux corrections (e.g., background removal, extinction correction). Flares A, B, C (Tab.~\ref{tab_nirx}) are still easy to compare, since they were observed by the same instruments in the same pass-bands, at roughly the same epoch. Flares H and I can also be added to this group, provided a background correction is applied and consistent extinctions employed\footnote{We used a $L'$ background of 3.6~mJy as given in caption of Fig.~5 and converted the extinctions quoted in caption of Tab.~1 \citep{hornstein07}.}.  Finally, peak NIR flux of the remaining seven flares (D, E, F, G, J, K, L) need to be extrapolated to the $L'$-band, assuming a particular infrared spectral index (see Sect.~\ref{color_section}), to ensure consistency with the other flares. 

As shown in Fig.~\ref{fig_nirx}, no particular correlation between the X-ray and NIR strengths is  apparent with the currently attainable statistics. Though the logarithmic scales of the plot axes do not reveal it, Flare~C really stands out of the group as a strong event in both X-rays and NIR.

\subsection{Sub-mm}

On April 1\st\ 2009, sub-mm observations overlapped with the NIR and X-ray coverages (Fig.~\ref{n1}). \sgra\ is found to be continuously variable at 870~\micron, though on a longer time scale than at shorter wavelengths\footnote{Analogous variability, on timescales ranging from hours to days, was also reported at wavelengths longward of $\sim$1~mm \cite[see e.g.,][]{wright93,bower02,zhao03,herrnstein04,miyazaki04,mauerhan05,yusef-zadeh07}.}.
We first note that there are no abrupt changes in the \apex\ light curve at the time the \vlt\ detects the rise of Flare~A. This means that a simultaneous sub-mm counterpart to this NIR flare is not detected as such, and corroborates similar past non-detections \citep{marrone08, yusef-zadeh08,eckart08b}. 

679$\pm$4~min after 2009--04--01 00:00:00 (UTC), the sub-mm flux of \sgra\ peaked at 5.4$\pm$0.2~Jy, and marks the summit of a dramatic sub-mm flare, among the brightest ever detected, that lasted for $\sim$150~min and that we will refer to as Flare~A'. When we subtract the background emission (taken as the minimum flux of the observation, 2.2~Jy), the peak is reduced to 3.2$\pm$0.2~mJy.
The rise time of Flare~A' is on the order of 20~min, which limits the size of the sub-mm flaring region to only $\sim$30\,$R_{\rm S}$ with the same argument as the one used in \textsection~\ref{nir_para}. Yet, this size constraint will probably be revised soon when Very Long Baseline Interferometry (VLBI) measurements is achieved in the sub-mm range. Recent size measurements at 1.3~mm have indeed already yielded an intriguing upper limit to the quiescent size of \sgra\ of $\sim$$40$~$\mu$as $\approx 4\,R_{\rm S}$ \citep{doeleman08}. 

Interestingly, Flare~A' lags Flare~A by $\sim$200~min and is not associated with any other flaring episode, at least as far as we can tell from the X-ray light curve. 
As summarized in Tab.~\ref{tab_delay}, this new flare detected with high confidence adds to a handful of past detections (sometimes yet debatable) of short mm/sub-mm activity in the aftermath of X-ray and/or NIR flares.
One way often proposed to relate these phenomena is that X-ray/NIR flares are 
{associated with a plasmoid expansion radiating }
 via synchrotron emission \citep[][see Paragraph~\ref{plasmoid_section}]{vanderlaan66}. If this plasmoid is also responsible for the production of X-rays through SSC emission, then as pointed out by \cite{marrone08}, there is a tight relation connecting the time lag between the X-ray/NIR peak fluxes ($F_{\rm X}$, $F_{\rm NIR}$) and the peak flux at a smaller frequency $\nu$. This lag is given by:

\begin{equation}
\Delta t_{\nu} \propto \left (\frac{F_{\rm X}}{F_{\rm NIR}}  B_0^{\,(p+2)/2} \nu^{-(4+p)/2} \right )^{1/\delta (3+2p)}
\end{equation}
 
\noindent where $B_0$ is the inital magnetic field, $p$ the spectral index of the radiating particles ($\mathrm{d}N/\mathrm{d}E\propto E^{-p}$), and $\delta$ is a parameter depicting the speed at which the radius $R$ of the plasmoid grows ($R\propto t^{\, \delta}$).
So far, only three flares were fully covered by contemporaneous X-ray, NIR, and sub-mm observations and allow for a test of this relation: Flares~A, H, and I. The comparison between H and I was done by \cite{marrone08}, who showed a consistency between the data and the plasmoid/SSC picture. Yet, when considering Flare~A/A', this tendency of ``the higher the ratio $F_{\rm X}/F_{\rm NIR}$, the longer the delay'' no longer holds. Indeed, Flare~I has a ratio about 10 times the ratio of Flare~A, and still presents a delay about twice shorter. This tends to indicate that X-ray/NIR flare and sub-mm flare emissions may not originate from the same particles (see also \textsection~\ref{plasmoid_section}).

Regarding the other two \apex\ observing runs of April 2009, no sub-mm flares were detected, notwithstanding an obvious flux decrease observed during the last run. The minimum flux reached was 2.2~Jy, justifying our previous choice for the steady quiescent background level.

\begin{table}

\renewcommand{\footnoterule}{}  

\centering
\begin{minipage}[t]{9cm}

\caption{Short mm and sub-mm flare delays relative to X-ray or NIR flares.}
\begin{tiny}
\label{tab_delay}

\begin{tabular}{lrcc}
\hline \hline
\vspace{0.05cm}

\textsc{References} \footnote{References: 
[1]~This work, 
[2]~\cite{marrone08},
[3]~\cite{yusef-zadeh08}, 
[4]~\cite{yusef-zadeh09}, 
[5]~\cite{eckart08b},
[6]~\cite{eckart06a}, 
[7]~\cite{eckart09}, 
[8]~\cite{yusef-zadeh06a}.  
}
&
\textsc{Date}       
& \textsc{Instruments}
& \textsc{Time lag} \footnote{Time by which the sub-mm peak lags the X-ray or NIR peak.}
\\

~ & ~ & ~ & [min] \\

\hline
[1] \dotfill						& 1 Apr 2009	& \apex~(870~\micron) 			& 200 \\

[2,3,4] \dotfill	 & 17 Jul 2006	& \sma~(1.3~mm)~/~\cso~(850~\micron) & 100 \\

[2] \dotfill 					& 31 Jul 2005	& \sma~(1.3~mm)				& 20 \\

[5] \dotfill 					& 3 Jun 2008	& \apex~(870~\micron) 			& 90 \\

[6,7] \dotfill		& 7 Jul 2004	& \sma~(890~\micron) 			& $<120$ \\

[4] \dotfill 				& 5 Apr 2007	& \sma~(1.3~mm) & 160 \\

[8] \dotfill 				& 3 Sep 2004	& \cso~(850~\micron) & 160 \\

\hline
\end{tabular}
\end{tiny}
\end{minipage}
\end{table}

\subsection{GeV $\gamma$-rays}
\label{gev}

The \fermi\ light curves in the top panels of Fig.~\ref{n1}, \ref{n2}, and \ref{n3}, are characterized by frequent data gaps on the order of 2~hr during which the GC is outside of the LAT's FOV.  
The statistics within each bin is {remarkably} poor; many 10~min bins do not contain any photon at all.
No significant peak in the light curves is detected.

Though Flare~A and most of Flare~A' fell outside the coverage of the GC by the LAT, the rising flank of Flare~B was coincident with the end of a LAT exposure. 
Zero photons were detected in the time bin at the peak of Flare~B.
Given the Poissonian statistics, the 3$\sigma$ (99.87\% confidence level) upper limit on the count rate of any source in the extraction region is 6.6 photons \citep{gehrels86} for the bin considered, which has an exposure of $6.2\times10^4$~s$^{-1}$\,cm$^{-2}$. This converts to a flux upper limit for Flare~B of $1.1\times10^{-4}$~\phperseccm, or $7.2\times10^{-8}$~\ergperseccm, and a luminosity upper limit of $5.5\times10^{38}$~\ergpersec\ (0.1--200~GeV) at 8~kpc.

{ This is not very constraining in regards of the flux quoted for the GC \fermi\ central point source, \sgrafermi, in the 1\st\ year catalogue\footnote{ These measurements were obtained with a modeling of the underlying diffuse emission.
A variability index of 18 was measured taking 11 time segments into account and implies that \sgrafermi\ is non-variable on a monthly timescale.} \citep{abdo10}: $1.7\times 10^{-6}$~\phperseccm\ (0.1--100~GeV) for a photon spectral index of $\sim$2.2.
}
\fermi\ thus confirms the presence of a GeV source in the GC, as once detected by the {Energetic Gamma-Ray ExperimenT (EGRET)} onboard the \textit{Compton Gamma-ray Observatory} \citep[][]{mayer-hasselwander98}.
The origin of \sgrafermi\ is unknown.
{It is quite possible that this GeV emission is produced in GC molecular clouds just as the TeV emission of the same region \citep{ballantyne07,melia10}.}
 But \sgra\ itself is also of course a valid candidate, given the positional coincidence of the LAT source and the supermassive black hole: \sgrafermi\ lies at the galactic coordinates [$l=359.941$\degree, $b=-0.051$\degree] with a 95\% confidence radius of $1.1'$, which is only $\sim$$21''$ away from \sgra.  
One way to unambiguously attribute this source to the supermassive black hole could have been the detection of a flaring activity correlated with the X-ray one. 
However, we did not detect any GeV signal with the sensitivity we achieved during Flare~B (which limited our detection capability to large GeV flares of amplitude at least $\sim$65 times the steady flux of \sgrafermi).
The present study is hence analogous to the unfruitful previous attempts to discover correlated emission between the X-ray flares of \sgra\ and the flux of the $\gamma$-ray sources \sgraigr\ and \sgrahess\ \citep[see][for the \integ\ and \hess\ non-detections, respectively]{trap10,aharonian08}.

\section{\boldmath NIR colors of \sgra}
\label{color_section}

 \begin{figure} 

   \centering
    \includegraphics[trim=2.3cm 8.5cm 2.5cm 8.5cm, clip=true, width=9cm]{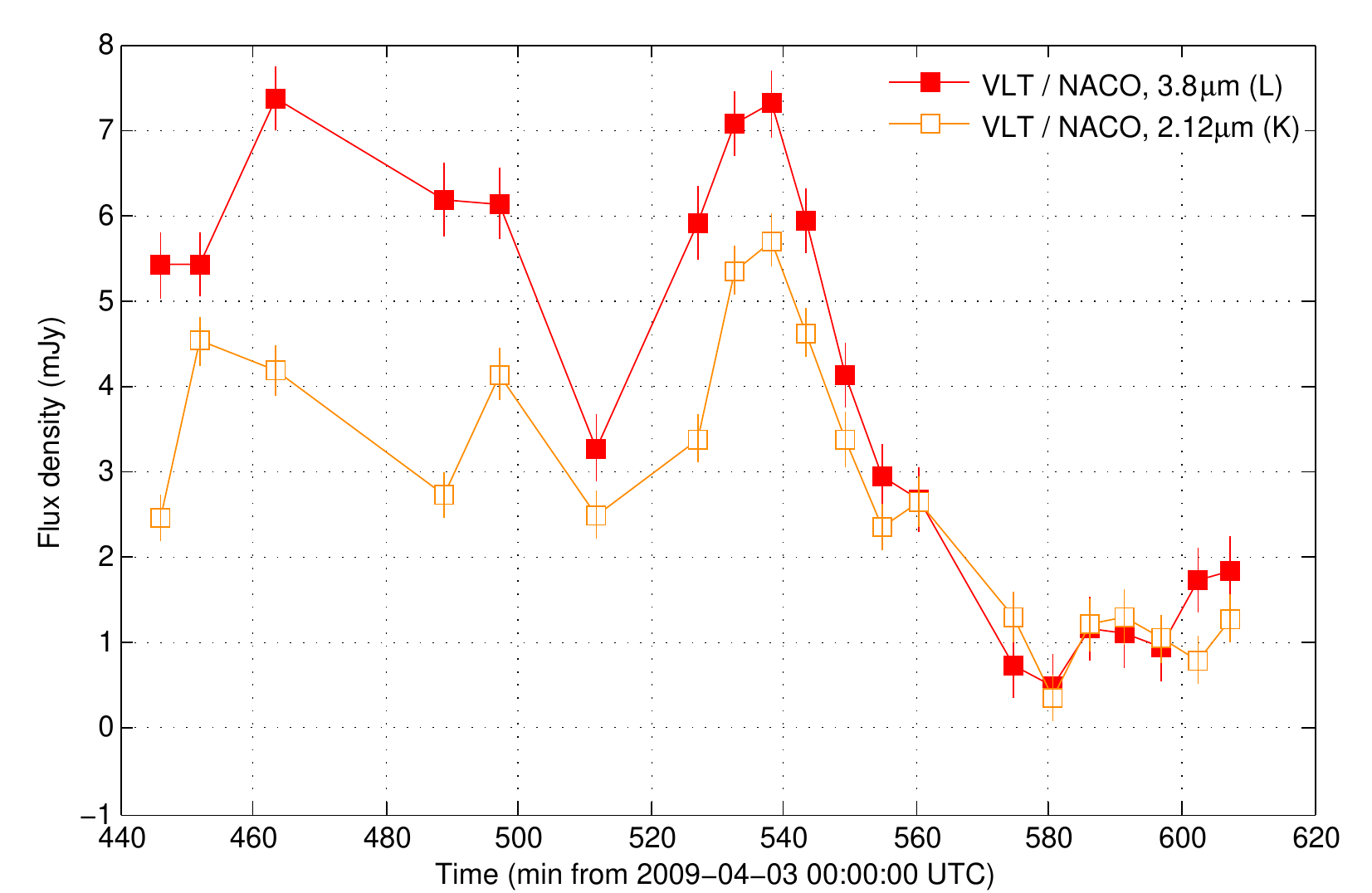}
    \includegraphics[trim=2.3cm 7cm 2.5cm 7cm, clip=true, width=9cm]{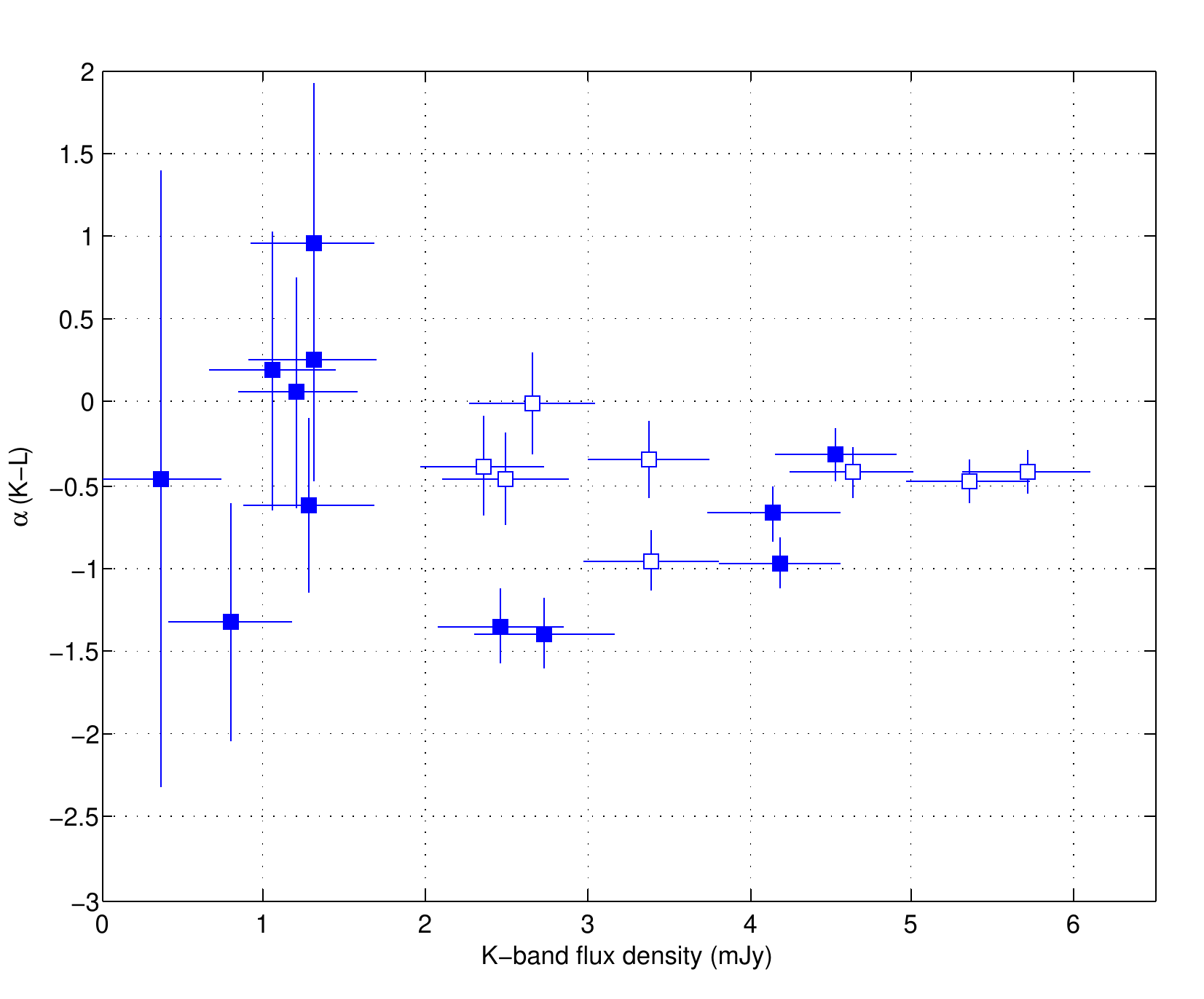}

   \caption{(\textit{Top panel}) NIR obervations of the 2\nd\ night background subtracted and corrected for extinction with $A_{L'}=1.8$~mag and $A_{K_{\rm s}}=A_{L'}+E(K_{\rm s}-L')=3.3$~mag. \textit{(Bottom panel}) Corresponding spectral index $\alpha$ plotted against dereddened $K_{\rm s}$-flux. Open squares indicate measurements made during Flare~B.}
   \label{color_katie}
\end{figure}

\begin{table*}

\renewcommand{\footnoterule}{}  
\begin{minipage}[t]{18cm}

\caption{Flare NIR spectral measurements.}
\centering
\begin{tiny}
\label{tab_color}

\begin{tabular}{lclrc}
\hline \hline
\vspace{0.05cm}

\textsc{References} \footnote{References: 
[1]~This work, 
[2]~\cite{hornstein07}, 
[3]~\cite{yusef-zadeh09}, 
[4]~\cite{dodds09},
[5]~\cite{trap10}, 
[6]~\cite{krabbe06}, 
[7]~\cite{gillessen06},  
[8]~\cite{eisenhauer05}, 
[9]~\cite{ghez05b}. }
&
\textsc{Dates}       
& \textsc{Instruments}
& \textsc{Bands}
& $\alpha$ \footnote{$\alpha$ is defined by $F_\nu \propto \nu^{\,\alpha}$. $^\dag$ indicates when there is a correlation observed between the flux and the spectral index. In the case of \cite{yusef-zadeh09}, only the brightest flares are reported.}
\\

~ & ~ & ~ & ~ & ~ \\

\hline
[1] \dotfill 						& 3 Apr 2009	& \vlt~/~NACO   & 2.1, 3.8~\micron 	& $-0.4\pm0.3$ \\

[2] \dotfill  					& 16, 31 Jul 2005, 2 May, 17 Jul 2006 	& \keck~/~NIRC2   & 1.6, 2.1, 3.8, 4.7~\micron 	& $-0.6\pm0.2$ \\

[3] \dotfill  				& 2, 5 Apr 2007	& \hst~/~NICMOS   & 1.45, 1.7~\micron 	& $-0.7\pm0.2$, $-1\pm0.3$$^\dag$ \\

[4,5] \dotfill 				& 4 Apr 2007	& \vlt~/~NACO, VISIR   & 3.8, 11.8~\micron 	& $>-1$ \\

[6] \dotfill  				 	& 29 Apr 2005	& \keck~/~OSIRIS   & 2.02--2.38~\micron 	& $-2.6\pm0.9$ \\

[7] \dotfill  					& 18 Jun 2005	& \vlt~/~SINFONI   & 2--2.45~\micron 	& $-$(0.6--$1.3)\pm0.2$$^\dag$ \\

[8] \dotfill 				& 15, 17 Jul 2004	& \vlt~/~SINFONI   & 1.7--2.45~\micron 	& $-4\pm1$ \\

[9] \dotfill  					& 26 Jul 2004	& \keck~/~NIRC2   & 2.1, 3.8~\micron 	& $-0.5\pm0.3$ \\

\hline
\end{tabular}
\end{tiny}
\end{minipage}
\end{table*}

Here we describe the results of the spectral imaging of \sgra\ we made during the 2\nd\ night of \vlt/NACO observations. 

To compute the NIR photon spectral index $\alpha$, defined by $F_{\nu}\propto \nu^{\,\alpha}$, we produced the $L'$- and $K_{\rm s}$-band light curves of \sgra\ presented in Fig.~\ref{color_katie} (top panel), for which we averaged the pairs of images taken at each filter switch. The error bars incorporate the standard deviation of these values as well as the error determined from the standard deviation of a comparison star of similar flux. Data points were interpolated to the same times and a large aperture was used in both bands to fully include the contribution of S\,17. A global background correction (including S\,17) was made by subtracting the lowest flux value. 
It is important to note that $\alpha$ is highly dependent on the excess color $E(K_{\rm s} - L')=A_{K_{\rm s}}-A_{L'}$, since:

\begin{equation}
\alpha = \frac{\log \left(10^{\,(A_{K_{\rm s}}-A_{L'})/2.5} \times F^{\rm \,red}_{K_{\rm s}}/F^{\rm \,red}_{L'}\right)}{\log \left(\nu_{K_{\rm s}}/\nu_{L'}\right)}
\end{equation}

\noindent where $F^{\rm \,red}$ are the reddened fluxes. 
Using known early-type stars\footnote{We used S\,2, S\,65, S\,66, S\,72, and S\,87 which have $K$ magnitudes between 13.5--14.6~mag and are spectroscopically identified to have spectral types between O9--B2 \citep{eisenhauer05}.} in the vicinity of \sgra, we determined an average color excess $E(K_{\rm s} - L')=1.5\pm0.1$ \citep[see e.g.][]{hornstein07}. 
 The corresponding values of $\alpha$ are plotted in Fig.~\ref{color_katie} (bottom panel) as a function of the $K_{\rm s}$-band flux. Given the statistics of the data, no particular correlation between $\alpha$ and the flux is found. The mean value obtained over the duration of Flare~B is  $\alpha = -0.4\pm0.3$.

For comparison, we list in Tab.~\ref{tab_color} all the previous attempts to measure the spectral indices of flares from \sgra. The different values of $\alpha$ thus obtained are however difficult to compare, since each of these works either probed different spectral ranges/flux amplitudes/epochs or used different instruments/techniques/reddening laws/backgrounds. Owing to the experimental setup used in this study, our result can only be easily compared to the work of \cite{hornstein07}. Likewise, these authors made use of a ground telescope with adaptive optics and the filter cycling technique, though they also accumulated data in the $H$- and $M_{\rm s}$-bands, and used \keck's laser guide star. They found a constant ``universal'' value for the index of several flares, no matter the instantaneous flux, of $\alpha=-0.6\pm0.2$, which is fully consistent with our result.
{ Such a stable index seems to support a sort of a IR ``quiescence'' on average even though this is debated \citep{dodds10bis}.}

It seems that all the values for $\alpha$ recently obtained suggest $\alpha \gtrsim -1$, which translates to $\beta \gtrsim 0$ (with $\nu F_\nu \propto \nu^{ \, \beta}$), thus justifying that ``infrared flares are blue'' and arise from a particle population distinct from the one producing the quiescent ``sub-mm bump'' \citep{dodds09,trap10}. Presuming a synchrotron origin for the NIR flares and the radiating electrons  distributed in a power-law of index $p=1-2\alpha$ ($\mathrm{d}N/\mathrm{d}E\propto E^{-p}$), this suggests that $p\lesssim3$, which has important implications for the fit of flare spectra and in particular the SSC components (see Paragraph~\ref{sed}).


\section{Discussion}
\label{discussion}

Below, we discuss a simple modeling of Flare~A/A' light curves and SED based on the expansion of a plasmoid releasing synchrotron and SSC radiation. Flare~B SED is then analyzed in the context of different non-thermal emission processes. Models for SEDs are essentially static and only meant to show in which direction our data may constrain the theory of \sgra's flares. Evolutionary effects and the physical conditions leading to such scenarios will have to be studied in more details in future works.

 \begin{figure*} 

   \centering
   \includegraphics[trim=0cm 6.5cm 0cm 6.5cm, clip=true, width=13cm]{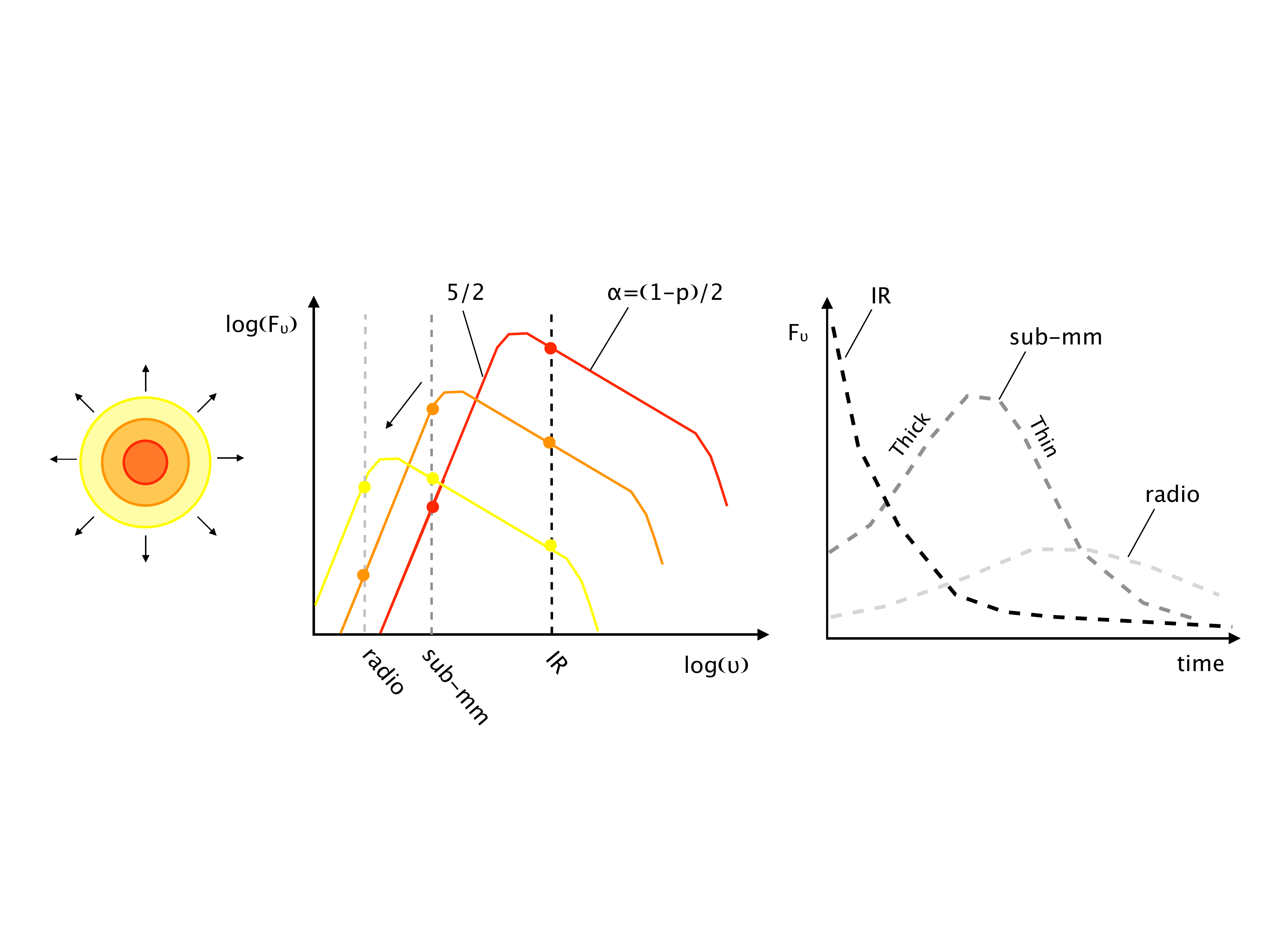}
   \caption{Sketches of the \citet{vanderlaan66} model. {As the plasmoid expands (\textit{left panel}), its initial self-absorbed synchrotron spectrum is shifted towards smaller frequencies (\textit{middle panel}). 
   If the source is observed at optically thin frequencies from the beginning,
then it remains optically thin and the flux keeps decreasing as time moves on (\textit{right panel}, NIR curve). Alternatively, if the 
source is initially optically thick, then the flux rises
until the source turns transparent and subsequently declines in the thin regime (\textit{right panel}, sub-mm and radio curves).}
  }
   \label{vanderlaan}
\end{figure*}

\subsection{Plasmoid expansion}
\label{plasmoid_section}

{ Over the past years, several authors have employed the adiabatic expansion of a plasmoid \citep{vanderlaan66} in the context of cm to sub-mm variability of \sgra, often observed to follow X-ray/NIR flares \citep[e.g.,][and the references listed in Tab.~\ref{tab_delay}]{yusef-zadeh06b,eckart06a}. The use of this expansion model for the flares of \sgra\ is motivated by:}
\begin{itemize}
\item the timelags separating X-ray/NIR flares from sub-mm flares (Tab.~\ref{tab_delay});
\item the timelags separating X-ray/NIR flares from radio flares \citep{yusef-zadeh09,kunneriath08};
\item the timelags between different radio bands within the same flare \citep{yusef-zadeh06b}\footnote{{Detailed models of these observations seem to exclude the simplest expansion scenario though, because the observed sizes of the radio emitting regions are bigger than van der Laan predicted sizes \citep{falcke09,maitra09}.}};
\item the increase of linear polarization as sub-mm flares reach their maximum \citep{marrone08};
\item the similarity between the flare timescales in X-ray/NIR and mm/sub-mm, which suggests that energy losses are not dominated by radiative cooling, with expansion a good alternative candidate\footnote{Radiative cooling effects may however also affect the expansion model in the mm range \citep{li09}.}.  
\end{itemize}

{ Fig.~\ref{vanderlaan} captures the essence of the model and gives the respective initial positions in the spectral domain of the wavebands used in this work. A ball of plasma of initial radius $R_0$, is filled with relavistic electrons of uniform volumic number density $n_{\rm e}$, arranged in a power-law ($\mathrm{d}n_{\rm e}/\mathrm{d}\gamma = k_{\rm e} \,\gamma^{-p}$, $\gamma_{\rm min} \leqslant \gamma \leqslant \gamma_{\rm max}$). 
This plasmon is pervaded by a magnetic field, $B$, and yields an initial synchrotron spectrum made of two connected power-laws (Fig.~\ref{vanderlaan}, middle panel, red curve). Below the initial synchrotron self-absorption (SSA) frequency $\nu_0$ \citep[$\propto \{R_0 \,k_{\rm e}\}^{\,2/(p+4)} B^{\,(p+2)/(p+4)}$,][]{gould79} at which the flux attains its maximal value, $F_0$, the spectrum is optically thick ($F_\nu \propto \nu^{\,5/2}$). For $\nu > \nu_0$, the spectrum is optically thin 
($F_\nu \propto \nu^{\,(1-p)/2}$) and breaks exponentially at a frequency proportional to $\gamma_{\rm max}^{\,2} B$. 
The dependence of the flux, $F_\nu$, on the radius, $R$, is given by:

\begin{equation}
\label{eq_flux}
 F_\nu = F_0 \left(\frac{\nu}{\nu_0}\right)^{5/2} \left(\frac{R}{R_0}\right)^{3} \frac{1-\exp(-\tau_\nu)}{1-\exp(-\tau_0)}
\end{equation}

\noindent where the optical depth, $\tau_\nu$, obeys the equation:

\begin{equation}
\label{eq_tau}
\tau_\nu = \tau_0 \left(\frac{\nu}{\nu_0}\right)^{-(p+4)/2} \left(\frac{R}{R_0}\right)^{-(2p+3)}
\end{equation}

\noindent and where the optical depth $\tau_0$ at $\nu_0$ is set by the 
condition that $F_\nu$ be maximal $\lbrack  \partial_\nu\,F_\nu (\nu_0,t)=0 \rbrack$, or:

\begin{equation}\label{tau0}
\exp(\tau_0) - \tau_0\,(p+4)/5-1=0~.
\end{equation}

\noindent In the following, we adopt a constant velocity, $V_{\rm exp}$,
for the expansion so that:

\begin{equation}
R(t) = R_0 + V_{\rm exp} (t-t_0)~. 
\end{equation}
}

To test this model with our multi-wavelength data of April~1, 
we propose two phenomenological scenarios (I and II) with
different sets of values for the parameters 
($p$, $R_0$, $\nu_0$, $F_0$, $V_{\rm exp}$, $t_0$, see Tab.~\ref{tab_param}). To assess the viability of these models
not only with the light curves but also spectrally, we calculate the plasmoid initial synchrotron 
spectra with particular values of $\gamma_{\rm min}$, $\gamma_{\rm max}$,
$B$, and $n_{\rm e}$, and also consider the naturally associated SSC emission, using the formalism of \cite{krawczynski04}\footnote{In the progress of this work, we found a problem in the treatment of the SSA
in H. Krawczynski's code, which derived from a typo in the numerical expression in SI units of the absorption coefficient given by \citet{longairII}, that should be written: $\chi_\nu = 20.9\, \kappa' \,B^{\,(p+2)/2}\,(5.67\times 10^8)^{\,p} \,b(p)\, \nu^{-(p+4)/2}$.}.
In all cases, the NIR emission is assumed to be optically thin, and so $\nu_{\rm submm} < \nu_0<\nu_{\rm NIR}$ (Fig.~\ref{vanderlaan}, middle panel). 
The NIR counterpart of \sgra\ is indeed known to be highly linearly polarized \citep{eckart06b, meyer06b,meyer06a,meyer07,trippe07} and, 
if it was optically thick, then an expansion model would
anyhow predict a sub-mm peak flux lower than the NIR one (Fig.~\ref{vanderlaan}, right panel), which is inconsistent with 
the observations.  
 \begin{figure} 

   \centering
      \includegraphics[trim=0cm 6.5cm 18cm 0cm, clip=true, width=9cm]{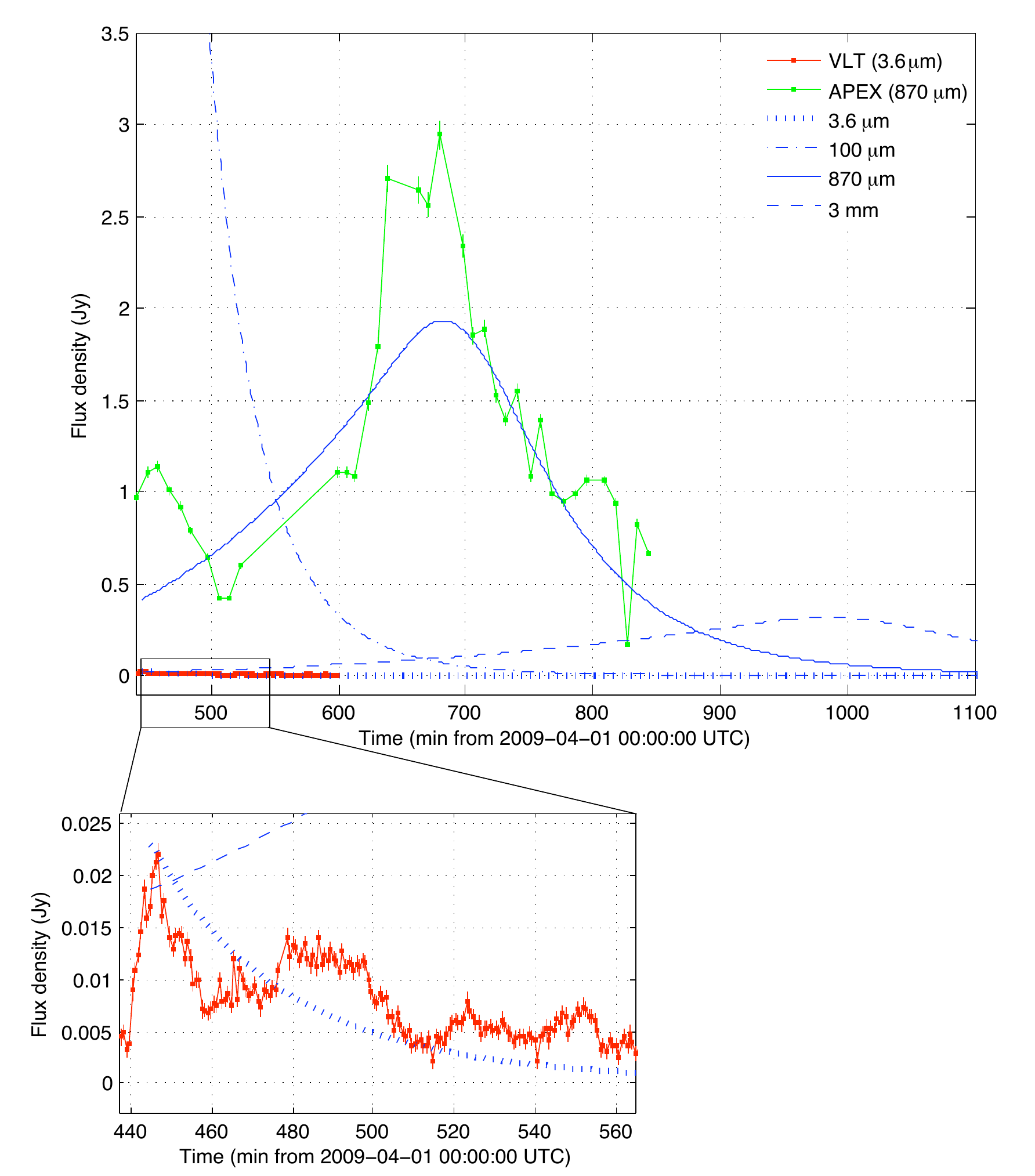}
         \includegraphics[trim=1.5cm 8cm 2.4cm 8cm, clip=true, width=9cm]{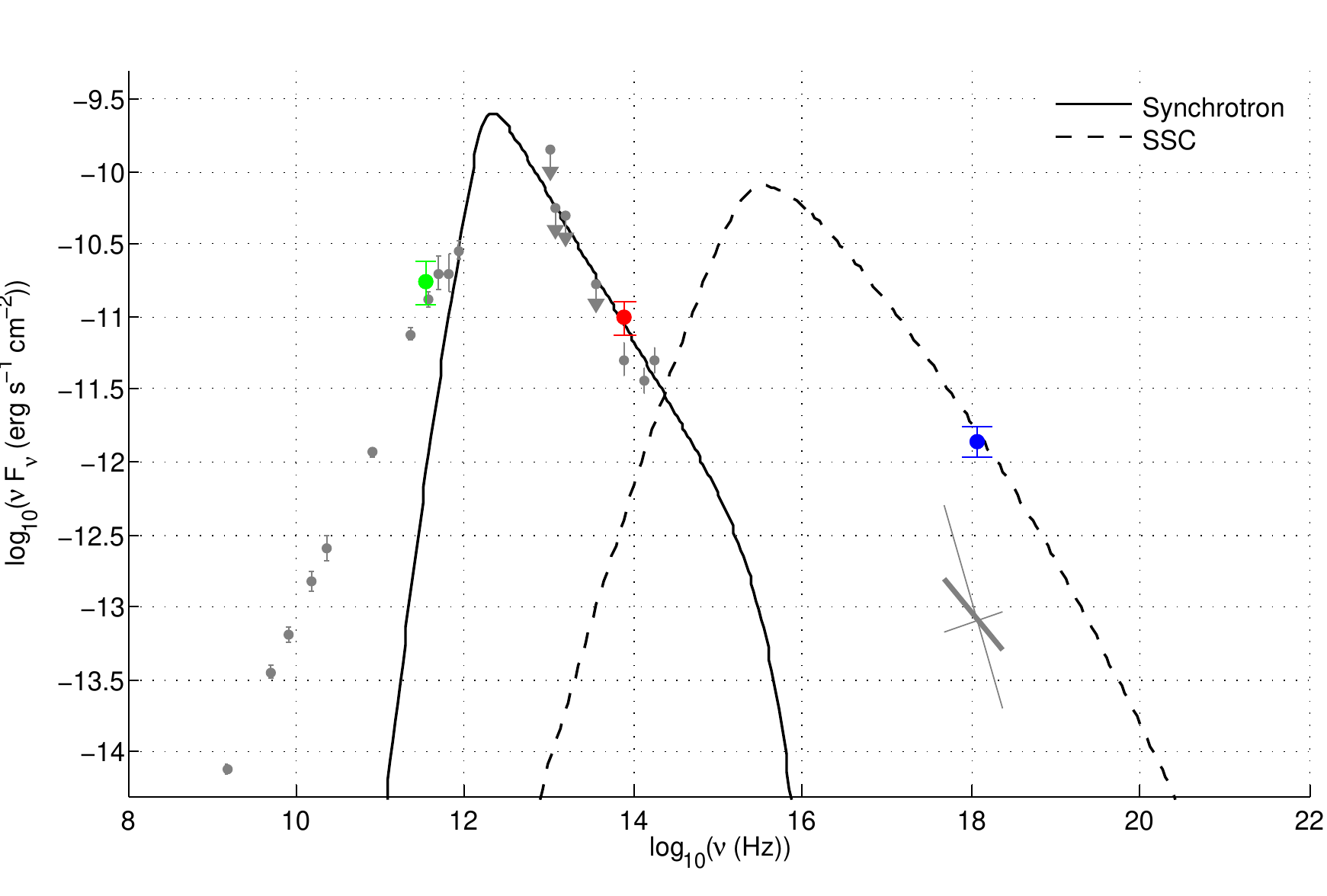}
         
  \caption{(\textit{Top panel}) \apex\ and \vlt\ light curves plotted on the same scale. Theoretical predictions of model~I for several wavelengths are overplotted in blue. (\textit{Middle panel}) Magnified view of the \vlt\ data. (\textit{Bottom panel}) The synchrotron and SSC curves give the initial SED of the plasmoid in model~I. They fit the simulaneous NIR and X-ray data (in red and blue respectively) of Flare~A. The non-simultaneous sub-mm data of Flare~A' are also indicated in green for information. 
 The references of the quiescence data in light gray are given in \cite{trap10}.}
   \label{model1}
\end{figure}

\textit{\textbf{Model I}}. 
The scenario that requires the least number of ingredients to accomodate the NIR, 
sub-mm, and X-ray light data all at once, is a plasmon initially producing the NIR and 
X-ray flare by synchrotron and SSC, respectively. As it then expands, the NIR/X-ray flare
decays and the sub-mm flare arises. Given the great difference between the NIR and
 sub-mm fluxes, this model demands an extremely steep power-law distribution ($p=5$) and a
 high maximum flux ($F_0=22$~Jy) at the SSA turnover. 
 We choose the different parameters so as to respect the flare size constraint ($R_0<10\,R_{\rm S}$) and the essential idea of the \citet{vanderlaan66} model ($\nu^{\rm sync}_{\rm min} <\nu_0$, where $\nu^{\rm sync}_{\rm min} \propto \gamma_{\rm min}^2 \nu_{\rm g}$, $\nu_{\rm g}=eB/\{2\pi\,m_{\rm e}\}$ is the gyration frequency, and $m_{\rm e}$ is the mass of the electron).
 Fig.~\ref{model1} (top and middle panels)
  shows that the overall NIR/sub-mm timescales and flux characteristics of 
  Flare~A/A' can be roughly reproduced {(predictions at 100~\micron\ and 3~mm are also plotted)}, and Fig.~\ref{model1} (bottom panel) that the observed initial SED constraints can be met.
However, the initial steep NIR spectrum contradicts the measurements of $\alpha$ made with Flare~B for example and exceeds the mid-infrared (MIR) upper limit obtained with \vlt/VISIR during the April 2007
  campaign \citep{trap10} by a factor $\sim$4. 
    We therefore disfavor model I, although we cannot categorically exclude it given our lack of infrared spectral information on Flare~A.

\begin{figure} 

   \centering
      \includegraphics[trim=1.5cm 7cm 2cm 7cm, clip=true, width=9cm]{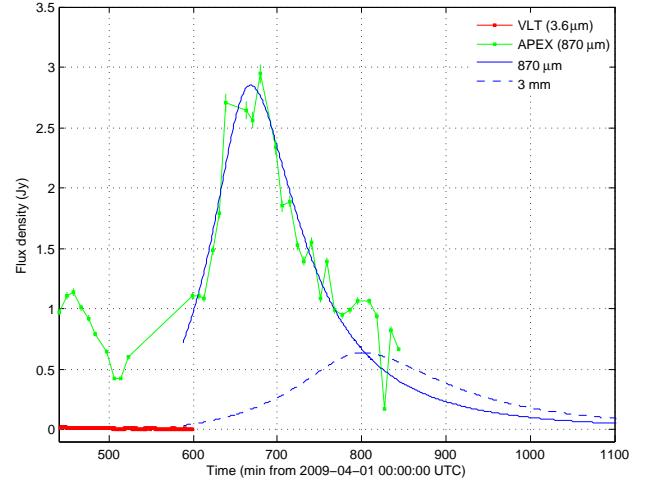}
   \caption{Same as top panel of Fig.~\ref{model1} for model~II.}
   \label{model2}
\end{figure}

\textit{\textbf{Model II}}. 
We now assume that Flare~A and A' are not directly connected by the same synchrotron power-law,
and that Flare~A simply acts as a trigger for a plasmoid expansion that will in term give
rise to Flare~A'. The plasmon power-law index can thus be chosen less steep ($p=2$) and the couple of parameters ($\gamma_{\rm max}$, $B$) so
that the initial plasmoid spectrum breaks in the far-infrared (FIR). It thus gives no significant synchrotron emission in the NIR and X-ray bands, and the same holds for the SSC radiation. 
Yet, if the expansion starts at the beginning of the NIR flare, the relatively long time interval  until the maximum of the sub-mm flare makes it difficult to obtain a narrow peak in the light curve at 870~\micron. 
Therefore, we allow the time at which the expansion 
is initiated to be postponed by 140~min, which gives an excellent fit to the \apex\ light curve (Fig.~\ref{model2}).
This is our favored model and it could be tested with radio millimeter data, since it predicts a distinctive 3~mm flare peaking at 0.6~Jy around 350~min after the onset of the NIR/X-ray flare.
We stress that, though ad hoc at this stage, this delayed expansion can be physically motivated in a ``coronal mass ejection'' picture by the fact that it would take the accelerated particles $\sim$200~min to be advected out of the inner parts of the accretion flow, where the high viscosity would prevent an immediate adiabatic expansion \citep[see][]{dodds10}. 

In all the above cases, we always used $R_0=R_{\rm S}$. But, as formulae \ref{eq_flux} and \ref{eq_tau} only involve the ratio $R/R_0$, we would have obtained the exact same sub-mm light curves by multiplying $R_0$ by any factor $x$, provided the velocity $V_{\rm exp}$ quoted in Tab.~\ref{tab_param} was multiplied by the same $x$. Since $10$\,$R_{\rm S}$ is an upper limit on the size of the flaring region, we find that the maximum expansion speed required is 0.05\,$c$, which is consistent with the other sub-relativistic 
expansions found previously \citep{yusef-zadeh08,eckart08b,eckart09}. This means, incidentally, that the plasmoid cannot escape the gravitational potential well of the black hole, unless it has a relativistic bulk motion, inside of a jet for example \citep{falcke09,maitra09}.

The modeling we present here compares well with previous works involving similar 
spectral windows \citep{eckart08b,eckart09}, although we used only one plasmon 
compared to the multi-component models needed in these works.
Nevertheless, our main result is that the simplest form of an abiabatic expansion scenario 
(model I) is inadequate to reproduce all our NIR/sub-mm/X-ray data at once, 
 so that we favor expansion models in which NIR/X-ray flares are not 
 initiated by the particles of the plasmon creating
the sub-mm flare. This is in contrast with the study of \cite{eckart08b}, 
in which models analogous to model I are generally favored. We note in particular that ``model BI''
of \cite{eckart08b} violates MIR upper limits on \sgra, and so does ``model AB$\alpha$$\phi3$'' 
in \cite{eckart09}, in spite of the NIR exponential break invoked to avoid the over-production
of NIR flux. 

We note also that a similar conclusion, disconnecting the NIR and radio portions of the spectrum in the frame of the \cite{vanderlaan66} model, was reached by \cite{mirabel98} in the study of discrete 
ejections from the microquasar GRS\,1915+105.

Expansion models for the sub-mm flares of \sgra\ should have two major telltale signatures: spectral indices in the sub-mm/mm range indicating an evolution from a thick to thin regime over the course of a flare, and high fluxes in the FIR { (100--300~\micron, Fig.~\ref{model1}, top panel)}.  
The upcoming observations of the Galactic nucleus with \textit{Herschel} and the \textit{Stratospheric Observatory for Infrared Astronomy} (\textit{SOFIA}) will perhaps allow the testing of these predictions.

\begin{table}

\renewcommand{\footnoterule}{}  
\begin{minipage}[t]{\columnwidth}

\caption{Parameters of Flare~A and Flare~B models.}
\centering
\begin{tiny}

\begin{tabular}{ccc|ccc}
\\
\hline 
\hline
\vspace{0.15cm}
~ & \multicolumn{2}{c}{\textsc{Flare A}} & \multicolumn{3}{c}{\textsc{Flare B}}\\
\textsc{Parameters~~~~} & \textsc{Model I} & \textsc{Model II} & \textsc{SSC1} & \textsc{SSC2} & \textsc{SB}\\
~ & ~ & ~ & ~ & ~ & ~\\
\hline
$p$\dotfill 						& 5 									& 2		& 2 					& 2.4							& 2.4--3.4\\
$R_0$ ($R_{\rm S}$)\dotfill 		&  1 								& 1		& 1					& 1							& 1\\
$\nu_0$ (Hz)\dotfill 				&  $2\times10^{12}$ 		 		& $2\times10^{12}$		&  --- 				&  ---		& --- \\
$F_0$ (Jy)\dotfill 				&  22 				 				& 17		&  --- 				&  ---		& ---\\
$V_{\rm exp}$ ($c$)\dotfill 			& 0.0017				   	 		& 0.005		&  --- 				&  ---		& ---\\
$t_0$ (min)\dotfill				& 445			  					& 588		&  --- 				&  ---		& ---\\
$\gamma_{\rm min}$\dotfill 		& 20										& 1			& 1					& 1							& 1\\
$\gamma_{\rm max}$\dotfill 		& $2\times10^3$		 			         & 100		& $1\times10^4$		& $2\times10^4$ 	   	 		& $6\times10^5$ \\
$B$ (G)\dotfill  					& 90  							& 100		& 2  				& 5							& 21\\
$n_{\rm e}$ (cm$^{-3}$)\dotfill		& $1\times10^{9}$  		& $3\times10^{9}$		& $1\times10^{9}$  	& $6\times10^{9}$			& $5\times10^{8}$\\
\hline
\\
\end{tabular}
\end{tiny}
\label{tab_param}
\end{minipage}
\end{table}

\subsection{SED modeling}
\label{sed}
 \begin{figure*} 

   \centering
 \includegraphics[trim=0cm 8.5cm 0cm 8.5cm, clip=true, width=12cm]{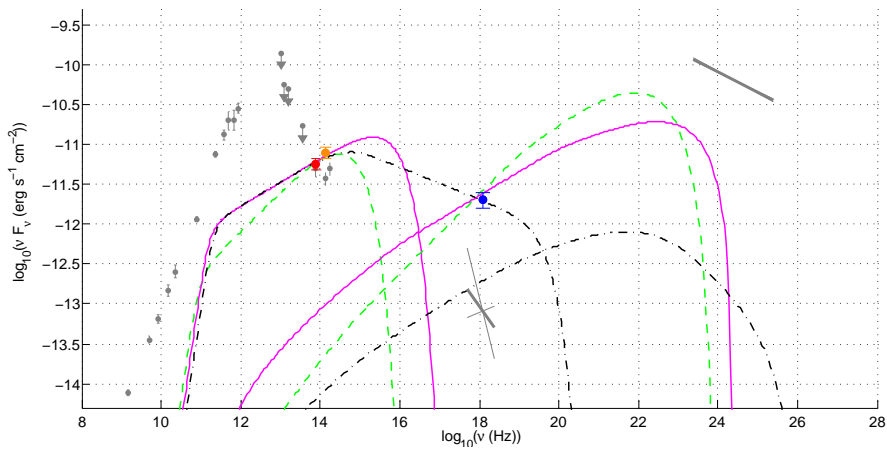}
 
   \caption{SED of Flare~B. $L'$-, $K_{\rm s}$-band, and X-ray peak fluxes are plotted in red, orange, and blue, respectively. NIR error bars account for the uncertainty in the extinction correction. The 1--100~GeV spectrum of \sgrafermi\ is indicated in dark gray. Models SSC1, SSC2, and SB correspond to the green dashed, magenta solid, black dash-dot lines, respectively.}
   \label{sed_color}
\end{figure*}

Though sub-mm and NIR flares may be unrelated, NIR and X-ray flares clearly derive 
from the same radiating particles given their simultaneity. 
{ Past theories of \sgra\ flares usually invoked a combination of synchrotron and inverse Compton mechanisms 
  \citep{markoff01,markoff07,liu02,liu04,liu06a,liu06b,yuan03,yuan04}.}
We analyze here synchrotron and SSC emission for Flare~B, for which we 
have NIR spectral information. 
{ This imposes a relatively hard electron spectrum ($p\lesssim3$, see Sect.~\ref{color_section}) contrary to model~I for Flare~A, which had a softer spectrum ($p=5$) as required by the expansion scenario.}
Since we do not have spectral information in X-rays, we propose three possible models (see Tab.~\ref{tab_param} and Fig.~\ref{sed_color}) to represent the SED of this flare, each having an X-ray slope either hard (SSC1), intermediate hard (SSC2), or soft (SB).   
In all three, the NIR flare is due to synchrotron emission. For SSC1 and SSC2, the X-ray flare results from SSC emission, which is the most commonly used configuration in the literature. 
Nonetheless, if the X-ray slope is soft, as for Flare~\#2 in \citet{porquet08}, then the SSC modeling is no longer valid, since it requires an enormous magnetic field strength of order 1000~G \citep{dodds09,trap10}. 
In this case, a broadband synchrotron spectrum with a break (SB) 
due to radiative cooling could give rise to a soft X-ray flare \citep[see][]{dodds09,dodds10,trap10}.

We note in Fig.~\ref{sed_color} that these flare spectra are all well below the 1--100 GeV spectrum of \sgrafermi, which is consistent with the non-detection reported in \textsection~\ref{gev}. 
In the progress of this work, we noticed a paper by \citet{kusunose10} modeling the time evolution of purely synchrotron (SB) scenarios for X-ray/NIR flares of \sgra. They predicted possible SSC counterparts observable with \fermi\ in a specific class of models, where the ratio of magnetic energy over particle energy is small. We have shown here that there was no detectable GeV excess during Flare~B but, we stress that given our low statistics, we could only constrain a putative GeV flare to a flux smaller than a few times $10^{-8}$~\ergperseccm, whereas their ``model~C'' induced a flux of only $\sim$$2\times10^{-11}$~\ergperseccm\ at maximum. Except for an extraordinarily bright event, it is very likely that such models will not be testable with \fermi\ data.


\section{Conclusions}

Only a handful of flaring events have been observed simultaneously accross a broad range of wavelengths up to now. Here, we have reported the detection of two new X-ray/NIR flares, a strong sub-mm outburst possibly associated with one of them, and their GeV $\gamma$-ray non-detection. 
These data allowed us to further characterize \sgra's variability and test non-thermal radiative mechanisms usually employed to describe the physics of the flares.

One flare was followed at the \vlt\ in two NIR spectral bands, thus demonstrating the feasibility of a filter switching technique to measure the color of \sgra\ with NACO. The resulting spectrum is found to be consistent with a power-law of index $\alpha=-0.4\pm0.3$ ($F_{\nu}\propto \nu^{\,\alpha}$), which is a ``blue'' spectrum ($\beta>0$ with $\nu F_\nu \propto \nu^{ \, \beta}$), implying that NIR flares arise from a particle population different from the one responsible for the quiescent sub-mm bump. 
For this flare, we could also check that there was no detectable in sync variation of the GeV $\gamma$-ray 
GC source.
The other NIR/X-ray flare was followed by a strong sub-mm flare, and was used to 
test the adiabatic expansion model which is often used to account for the
longer wavelength flare delays. In this context, we find it difficult to relate the NIR and sub-mm events by the same single synchrotron plasmoid, as it would induce a too steep infrared spectrum in regards of other flares. 
Even assuming the NIR/X-ray event just acted as a trigger for the plasmon expansion, 
the long delay observed between the NIR/X-ray and sub-mm peaks spreads the theoretical light curve of the sub-mm flare too much to reproduce the data.
However, we suggest an expanding plasmon could still drive the sub-mm flare light curve provided it started expanding about a couple of hours after the X-ray/NIR event. 
These models can all be tested by radio observations obtained in April 2009 and predict a significant flux in the FIR that could be observed by the \textit{Herschel} and \textit{SOFIA} observatories in the near future. Hopefully, the source expansion scenario will also soon be put to test by VLBI direct size measurements in the sub-mm range \citep{doeleman09}.   
These studies of \sgra's present variability and the ones pursuing giant outbursts experienced by \sgra\ in a recent past \citep{terrier10,ponti10}, should thus gradually bring an overall characterization of our Galactic nucleus' activity and help us put it into the general perspective of AGNs.


\begin{acknowledgements}

GT and KDE warmly thank C.~Lidman and G.~Carraro for assistance with the NIR observations at Paranal, and the observers at the Chajnantor site (C.~De Breuck, A.~Lundgren, M.~Dumke) for their efforts in planning and making the \apex~obs.
GT is also grateful to V.~Reveret, F.~Mattana, M.~Sakano, and D.~Marrone for enlightening discussions.
This work has been partly supported by the french Agence Nationale pour la Recherche through grant ANR--06--JC--0047.
GP acknowledges support via an EU Marie Curie Intra-European
Fellowship under contract FP7--PEOPLE--2009--IEF--254279.
The \xmm~project is an ESA Science Mission with instruments and contributions directly funded by ESA Member States and the USA (NASA).     
Parts of this study are based on observations made with ESO Telescopes at Paranal and Chajnantor Observatories under the programs LP 183.B-0100 and 69.A-0123. 
\end{acknowledgements}


\bibliographystyle{aa}

\bibliography{gc_trap.bib}

\begin{thebibliography}{92}
\expandafter\ifx\csname natexlab\endcsname\relax\def\natexlab#1{#1}\fi

\bibitem[{{Abdo} {et~al.}(2010){Abdo}, {Ackermann}, {Ajello}, {Allafort},
  {Atwood}, {Baldini}, {Ballet}, {Barbiellini}, {Baring}, {Bastieri},
  {Baughman}, {Bechtol}, {Bellazzini}, {Berenji}, {Blandford}, {Bloom},
  {Bonamente}, {Borgland}, {Bouvier}, {Bregeon}, {Brez}, {Brigida}, {Bruel},
  {Burnett}, {Buson}, {Caliandro}, {Cameron}, {Caraveo}, {Carrigan},
  {Casandjian}, {Cecchi}, {{\c C}elik}, {Chekhtman}, {Cheung}, {Chiang},
  {Ciprini}, {Claus}, {Cohen-Tanugi}, {Conrad}, {Dermer}, {de Luca}, {de
  Palma}, {Dormody}, {Silva}, {Drell}, {Dubois}, {Dumora}, {Farnier},
  {Favuzzi}, {Fegan}, {Focke}, {Fortin}, {Frailis}, {Fukazawa}, {Funk},
  {Fusco}, {Gargano}, {Gasparrini}, {Gehrels}, {Germani}, {Giavitto},
  {Giebels}, {Giglietto}, {Giordano}, {Glanzman}, {Godfrey}, {Grenier},
  {Grondin}, {Grove}, {Guillemot}, {Guiriec}, {Hadasch}, {Harding}, {Hays},
  {Hobbs}, {Horan}, {Hughes}, {Jackson}, {J{\'o}hannesson}, {Johnson},
  {Johnson}, {Johnson}, {Kamae}, {Katagiri}, {Kataoka}, {Kawai}, {Kerr},
  {Kn{\"o}dlseder}, {Kuss}, {Lande}, {Latronico}, {Lee}, {Lemoine-Goumard},
  {Llena Garde}, {Longo}, {Loparco}, {Lott}, {Lovellette}, {Lubrano}, {Makeev},
  {Manchester}, {Marelli}, {Mazziotta}, {McConville}, {McEnery}, {McGlynn},
  {Meurer}, {Michelson}, {Mitthumsiri}, {Mizuno}, {Moiseev}, {Monte},
  {Monzani}, {Morselli}, {Moskalenko}, {Murgia}, {Nakamori}, {Nolan}, {Norris},
  {Noutsos}, {Nuss}, {Ohsugi}, {Omodei}, {Orlando}, {Ormes}, {Ozaki},
  {Paneque}, {Panetta}, {Parent}, {Pelassa}, {Pepe}, {Pesce-Rollins},
  {Pierbattista}, {Piron}, {Porter}, {Rain{\`o}}, {Rando}, {Ray}, {Razzano},
  {Reimer}, {Reimer}, {Reposeur}, {Ritz}, {Rochester}, {Rodriguez}, {Romani},
  {Roth}, {Ryde}, {Sadrozinski}, {Sander}, {Saz Parkinson}, {Sgr{\`o}},
  {Siskind}, {Smith}, {Smith}, {Spandre}, {Spinelli}, {Starck}, {Strickman},
  {Suson}, {Takahashi}, {Takahashi}, {Tanaka}, {Thayer}, {Thayer}, {Thompson},
  {Tibaldo}, {Torres}, {Tosti}, {Tramacere}, {Usher}, {Van Etten}, {Vasileiou},
  {Venter}, {Vilchez}, {Vitale}, {Waite}, {Wang}, {Watters}, {Weltevrede},
  {Winer}, {Wood}, {Ylinen}, \& {Ziegler}}]{abdo10}
{Abdo}, A.~A., {Ackermann}, M., {Ajello}, M., {et~al.} 2010, ApJ Submitted

\bibitem[{{Aharonian} {et~al.}(2004){Aharonian}, {Akhperjanian}, {Aye},
  {Bazer-Bachi}, {Beilicke}, {Benbow}, {Berge}, {Berghaus}, {Bernl{\"o}hr},
  {Boisson}, {Bolz}, {Borgmeier}, {Braun}, {Breitling}, {Brown}, {Bussons
  Gordo}, {Chadwick}, {Chounet}, {Cornils}, {Costamante}, {Degrange},
  {Djannati-Ata{\"\i}}, {O'C.~Drury}, {Dubus}, {Ergin}, {Espigat}, {Feinstein},
  {Fleury}, {Fontaine}, {Funk}, {Gallant}, {Giebels}, {Gillessen}, {Goret},
  {Hadjichristidis}, {Hauser}, {Heinzelmann}, {Henri}, {Hermann}, {Hinton},
  {Hofmann}, {Holleran}, {Horns}, {de Jager}, {Jung}, {Kh{\'e}lifi}, {Komin},
  {Konopelko}, {Latham}, {Le Gallou}, {Lemi{\`e}re}, {Lemoine}, {Leroy},
  {Lohse}, {Marcowith}, {Masterson}, {McComb}, {de Naurois}, {Nolan},
  {Noutsos}, {Orford}, {Osborne}, {Ouchrif}, {Panter}, {Pelletier}, {Pita},
  {P{\"u}hlhofer}, {Punch}, {Raubenheimer}, {Raue}, {Raux}, {Rayner},
  {Redondo}, {Reimer}, {Reimer}, {Ripken}, {Rob}, {Rolland}, {Rowell},
  {Sahakian}, {Saug{\'e}}, {Schlenker}, {Schlickeiser}, {Schuster}, {Schwanke},
  {Siewert}, {Sol}, {Steenkamp}, {Stegmann}, {Tavernet}, {Terrier},
  {Th{\'e}oret}, {Tluczykont}, {Vasileiadis}, {Venter}, {Vincent}, {Visser},
  {V{\"o}lk}, \& {Wagner}}]{aharonian04}
{Aharonian}, F., {Akhperjanian}, A.~G., {Aye}, K.-M., {et~al.} 2004, A\&A, 425,
  L13

\bibitem[{{Aharonian} {et~al.}(2008){Aharonian}, {Akhperjanian}, {Barres de
  Almeida}, {Bazer-Bachi}, {Becherini}, {Behera}, {Benbow}, {Bernl{\"o}hr},
  {Boisson}, {Bochow}, {Borrel}, {Braun}, {Brion}, {Brucker}, {Brun},
  {B{\"u}hler}, {Bulik}, {B{\"u}sching}, {Boutelier}, {Carrigan}, {Chadwick},
  {Charbonnier}, {Chaves}, {Cheesebrough}, {Chounet}, {Clapson}, {Coignet},
  {Dalton}, {Degrange}, {Deil}, {Dickinson}, {Djannati-Ata{\"i}}, {Domainko},
  {O'C.~Drury}, {Dubois}, {Dubus}, {Dyks}, {Dyrda}, {Egberts},
  {Emmanoulopoulos}, {Espigat}, {Farnier}, {Feinstein}, {Fiasson},
  {F{\"o}rster}, {Fontaine}, {F{\"u}{\ss}ling}, {Gabici}, {Gallant},
  {G{\'e}rard}, {Giebels}, {Glicenstein}, {Gl{\"u}ck}, {Goret},
  {Hadjichristidis}, {Hauser}, {Hauser}, {Heinz}, {Heinzelmann}, {Henri},
  {Hermann}, {Hinton}, {Hoffmann}, {Hofmann}, {Holleran}, {Hoppe}, {Horns},
  {Jacholkowska}, {de Jager}, {Jung}, {Katarzy{\'n}ski}, {Kaufmann},
  {Kendziorra}, {Kerschhaggl}, {Khangulyan}, {Kh{\'e}lifi}, {Keogh}, {Komin},
  {Kosack}, {Lamanna}, {Lenain}, {Lohse}, {Marandon}, {Martin},
  {Martineau-Huynh}, {Marcowith}, {Maurin}, {McComb}, {Medina}, {Moderski},
  {Moulin}, {Naumann-Godo}, {de Naurois}, {Nedbal}, {Nekrassov}, {Niemiec},
  {Nolan}, {Ohm}, {Olive}, {de O{\~n}a Wilhelmi}, {Orford}, {Osborne},
  {Ostrowski}, {Panter}, {Pedaletti}, {Pelletier}, {Petrucci}, {Pita},
  {P{\"u}hlhofer}, {Punch}, {Quirrenbach}, {Raubenheimer}, {Raue}, {Rayner},
  {Renaud}, {Rieger}, {Ripken}, {Rob}, {Rosier-Lees}, {Rowell}, {Rudak},
  {Rulten}, {Ruppel}, {Sahakian}, {Santangelo}, {Schlickeiser}, {Sch{\"o}ck},
  {Schr{\"o}der}, {Schwanke}, {Schwarzburg}, {Schwemmer}, {Shalchi}, {Skilton},
  {Sol}, {Spangler}, {Stawarz}, {Steenkamp}, {Stegmann}, {Superina}, {Tam},
  {Tavernet}, {Terrier}, {Tibolla}, {van Eldik}, {Vasileiadis}, {Venter},
  {Vialle}, {Vincent}, {Vivier}, {V{\"o}lk}, {Volpe}, {Wagner}, {Ward},
  {Zdziarski}, \& {Zech}}]{aharonian08}
{Aharonian}, F., {Akhperjanian}, A.~G., {Barres de Almeida}, U., {et~al.} 2008,
  \aap, 492, L25

\bibitem[{{Atwood} {et~al.}(2009){Atwood}, {Abdo}, {Ackermann}, {Althouse},
  {Anderson}, {Axelsson}, {Baldini}, {Ballet}, {Band}, {Barbiellini},
  {Bartelt}, {Bastieri}, {Baughman}, {Bechtol}, {B{\'e}d{\'e}r{\`e}de},
  {Bellardi}, {Bellazzini}, {Berenji}, {Bignami}, {Bisello}, {Bissaldi},
  {Blandford}, {Bloom}, {Bogart}, {Bonamente}, {Bonnell}, {Borgland},
  {Bouvier}, {Bregeon}, {Brez}, {Brigida}, {Bruel}, {Burnett}, {Busetto},
  {Caliandro}, {Cameron}, {Caraveo}, {Carius}, {Carlson}, {Casandjian},
  {Cavazzuti}, {Ceccanti}, {Cecchi}, {Charles}, {Chekhtman}, {Cheung},
  {Chiang}, {Chipaux}, {Cillis}, {Ciprini}, {Claus}, {Cohen-Tanugi},
  {Condamoor}, {Conrad}, {Corbet}, {Corucci}, {Costamante}, {Cutini}, {Davis},
  {Decotigny}, {DeKlotz}, {Dermer}, {de Angelis}, {Digel}, {do Couto e Silva},
  {Drell}, {Dubois}, {Dumora}, {Edmonds}, {Fabiani}, {Farnier}, {Favuzzi},
  {Flath}, {Fleury}, {Focke}, {Funk}, {Fusco}, {Gargano}, {Gasparrini},
  {Gehrels}, {Gentit}, {Germani}, {Giebels}, {Giglietto}, {Giommi}, {Giordano},
  {Glanzman}, {Godfrey}, {Grenier}, {Grondin}, {Grove}, {Guillemot}, {Guiriec},
  {Haller}, {Harding}, {Hart}, {Hays}, {Healey}, {Hirayama}, {Hjalmarsdotter},
  {Horn}, {Hughes}, {J{\'o}hannesson}, {Johansson}, {Johnson}, {Johnson},
  {Johnson}, {Johnson}, {Kamae}, {Katagiri}, {Kataoka}, {Kavelaars}, {Kawai},
  {Kelly}, {Kerr}, {Klamra}, {Kn{\"o}dlseder}, {Kocian}, {Komin}, {Kuehn},
  {Kuss}, {Landriu}, {Latronico}, {Lee}, {Lee}, {Lemoine-Goumard}, {Lionetto},
  {Longo}, {Loparco}, {Lott}, {Lovellette}, {Lubrano}, {Madejski}, {Makeev},
  {Marangelli}, {Massai}, {Mazziotta}, {McEnery}, {Menon}, {Meurer},
  {Michelson}, {Minuti}, {Mirizzi}, {Mitthumsiri}, {Mizuno}, {Moiseev},
  {Monte}, {Monzani}, {Moretti}, {Morselli}, {Moskalenko}, {Murgia},
  {Nakamori}, {Nishino}, {Nolan}, {Norris}, {Nuss}, {Ohno}, {Ohsugi}, {Omodei},
  {Orlando}, {Ormes}, {Paccagnella}, {Paneque}, {Panetta}, {Parent}, {Pearce},
  {Pepe}, {Perazzo}, {Pesce-Rollins}, {Picozza}, {Pieri}, {Pinchera}, {Piron},
  {Porter}, {Poupard}, {Rain{\`o}}, {Rando}, {Rapposelli}, {Razzano}, {Reimer},
  {Reimer}, {Reposeur}, {Reyes}, {Ritz}, {Rochester}, {Rodriguez}, {Romani},
  {Roth}, {Russell}, {Ryde}, {Sabatini}, {Sadrozinski}, {Sanchez}, {Sander},
  {Sapozhnikov}, {Parkinson}, {Scargle}, {Schalk}, {Scolieri}, {Sgr{\`o}},
  {Share}, {Shaw}, {Shimokawabe}, {Shrader}, {Sierpowska-Bartosik}, {Siskind},
  {Smith}, {Smith}, {Spandre}, {Spinelli}, {Starck}, {Stephens}, {Strickman},
  {Strong}, {Suson}, {Tajima}, {Takahashi}, {Takahashi}, {Tanaka}, {Tenze},
  {Tether}, {Thayer}, {Thayer}, {Thompson}, {Tibaldo}, {Tibolla}, {Torres},
  {Tosti}, {Tramacere}, {Turri}, {Usher}, {Vilchez}, {Vitale}, {Wang},
  {Watters}, {Winer}, {Wood}, {Ylinen}, \& {Ziegler}}]{atwood09}
{Atwood}, W.~B., {Abdo}, A.~A., {Ackermann}, M., {et~al.} 2009, \apj, 697, 1071

\bibitem[{{Baganoff} {et~al.}(2003){Baganoff}, {Maeda}, {Morris}, {Bautz},
  {Brandt}, {Cui}, {Doty}, {Feigelson}, {Garmire}, {Pravdo}, {Ricker}, \&
  {Townsley}}]{baganoff03a}
{Baganoff}, F., {Maeda}, Y., {Morris}, M., {et~al.} 2003, \apj\, 591, 891

\bibitem[{{Baganoff} {et~al.}(2001){Baganoff}, {Bautz}, {Brandt}, {Chartas},
  {Feigelson}, {Garmire}, {Maeda}, {Morris}, {Ricker}, {Townsley}, \&
  {Walter}}]{baganoff01}
{Baganoff}, F.~K., {Bautz}, M.~W., {Brandt}, W.~N., {et~al.} 2001, Nature, 413,
  45

\bibitem[{{Balick} \& {Brown}(1974)}]{balick74}
{Balick}, B. \& {Brown}, R. 1974, ApJ, 194, 265

\bibitem[{{Ballantyne} {et~al.}(2007){Ballantyne}, {Melia}, {Liu}, \&
  {Crocker}}]{ballantyne07}
{Ballantyne}, D.~R., {Melia}, F., {Liu}, S., \& {Crocker}, R.~M. 2007, \apjl,
  657, L13

\bibitem[{{B{\'e}langer} {et~al.}(2004){B{\'e}langer}, {Goldwurm}, {Goldoni},
  {Paul}, {Terrier}, {Falanga}, {Ubertini}, {Bazzano}, {Del Santo}, {Winkler},
  {Parmar}, {Kuulkers}, {Ebisawa}, {Roques}, {Lund}, \& {Melia}}]{belanger04}
{B{\'e}langer}, G., {Goldwurm}, A., {Goldoni}, P., {et~al.} 2004, ApJ, 601,
  L163

\bibitem[{{B{\'e}langer} {et~al.}(2005){B{\'e}langer}, {Goldwurm}, {Melia},
  {Ferrando}, {Grosso}, {Porquet}, {Warwick}, \& {Yusef-Zadeh}}]{belanger05}
{B{\'e}langer}, G., {Goldwurm}, A., {Melia}, F., {et~al.} 2005, ApJ, 635, 1095

\bibitem[{{Bower} {et~al.}(2002){Bower}, {Falcke}, {Sault}, \&
  {Backer}}]{bower02}
{Bower}, G.~C., {Falcke}, H., {Sault}, R.~J., \& {Backer}, D.~C. 2002, \apj,
  571, 843

\bibitem[{{Dodds-Eden} {et~al.}(2010{\natexlab{a}}){Dodds-Eden}, {Gillessen},
  {Fritz}, {Eisenhauer}, {Trippe}, {Genzel}, {Ott}, {Bartko}, {Pfuhl}, {Bower},
  {Goldwurm}, {Porquet}, {Trap}, \& {Yusef-Zadeh}}]{dodds10bis}
{Dodds-Eden}, K., {Gillessen}, S., {Fritz}, T.~K., {et~al.} 2010{\natexlab{a}},
  ArXiv:1008.1984

\bibitem[{{Dodds-Eden} {et~al.}(2009){Dodds-Eden}, {Porquet}, {Trap},
  {Quataert}, {Haubois}, {Gillessen}, {Grosso}, {Pantin}, {Falcke}, {Rouan},
  {Genzel}, {Hasinger}, {Goldwurm}, {Yusef-Zadeh}, {Clenet}, {Trippe},
  {Lagage}, {Bartko}, {Eisenhauer}, {Ott}, {Paumard}, {Perrin}, {Yuan},
  {Fritz}, \& {Mascetti}}]{dodds09}
{Dodds-Eden}, K., {Porquet}, D., {Trap}, G., {et~al.} 2009, \apj, 698, 676

\bibitem[{{Dodds-Eden} {et~al.}(2010{\natexlab{b}}){Dodds-Eden}, {Sharma},
  {Quataert}, {Genzel}, {Gillessen}, {Eisenhauer}, \& {Porquet}}]{dodds10}
{Dodds-Eden}, K., {Sharma}, P., {Quataert}, E., {et~al.} 2010{\natexlab{b}},
  ArXiv:1005.0389

\bibitem[{{Doeleman} {et~al.}(2009){Doeleman}, {Fish}, {Broderick}, {Loeb}, \&
  {Rogers}}]{doeleman09}
{Doeleman}, S.~S., {Fish}, V.~L., {Broderick}, A.~E., {Loeb}, A., \& {Rogers},
  A.~E.~E. 2009, \apj, 695, 59

\bibitem[{{Doeleman} {et~al.}(2008){Doeleman}, {Weintroub}, {Rogers},
  {Plambeck}, {Freund}, {Tilanus}, {Friberg}, {Ziurys}, {Moran}, {Corey},
  {Young}, {Smythe}, {Titus}, {Marrone}, {Cappallo}, {Bock}, {Bower},
  {Chamberlin}, {Davis}, {Krichbaum}, {Lamb}, {Maness}, {Niell}, {Roy},
  {Strittmatter}, {Werthimer}, {Whitney}, \& {Woody}}]{doeleman08}
{Doeleman}, S.~S., {Weintroub}, J., {Rogers}, A.~E.~E., {et~al.} 2008, Nature,
  455, 78

\bibitem[{{Eckart} {et~al.}(2004){Eckart}, {Baganoff}, {Morris}, {Bautz},
  {Brandt}, {Garmire}, {Genzel}, {Ott}, {Ricker}, {Straubmeier}, {Viehmann},
  {Sch{\"o}del}, {Bower}, \& {Goldston}}]{eckart04}
{Eckart}, A., {Baganoff}, F.~K., {Morris}, M., {et~al.} 2004, A\&A, 427, 1

\bibitem[{{Eckart} {et~al.}(2009){Eckart}, {Baganoff}, {Morris}, {Kunneriath},
  {Zamaninasab}, {Witzel}, {Sch{\"o}del}, {Garc{\'{\i}}a-Mar{\'{\i}}n},
  {Meyer}, {Bower}, {Marrone}, {Bautz}, {Brandt}, {Garmire}, {Ricker},
  {Straubmeier}, {Roberts}, {Muzic}, {Mauerhan}, \& {Zensus}}]{eckart09}
{Eckart}, A., {Baganoff}, F.~K., {Morris}, M.~R., {et~al.} 2009, \aap, 500, 935

\bibitem[{{Eckart} {et~al.}(2006{\natexlab{a}}){Eckart}, {Baganoff},
  {Sch{\"o}del}, {Morris}, {Genzel}, {Bower}, {Marrone}, {Moran}, {Viehmann},
  {Bautz}, {Brandt}, {Garmire}, {Ott}, {Trippe}, {Ricker}, {Straubmeier},
  {Roberts}, {Yusef-Zadeh}, {Zhao}, \& {Rao}}]{eckart06a}
{Eckart}, A., {Baganoff}, F.~K., {Sch{\"o}del}, R., {et~al.}
  2006{\natexlab{a}}, A\&A, 450, 535

\bibitem[{{Eckart} {et~al.}(2008{\natexlab{a}}){Eckart}, {Baganoff},
  {Zamaninasab}, {Morris}, {Sch{\"o}del}, {Meyer}, {Muzic}, {Bautz}, {Brandt},
  {Garmire}, {Ricker}, {Kunneriath}, {Straubmeier}, {Duschl}, {Dovciak},
  {Karas}, {Markoff}, {Najarro}, {Mauerhan}, {Moultaka}, \&
  {Zensus}}]{eckart08a}
{Eckart}, A., {Baganoff}, F.~K., {Zamaninasab}, M., {et~al.}
  2008{\natexlab{a}}, A\&A, 479, 625

\bibitem[{{Eckart} {et~al.}(2008{\natexlab{b}}){Eckart}, {Sch{\"o}del},
  {Garc{\'{\i}}a-Mar{\'{\i}}n}, {Witzel}, {Weiss}, {Baganoff}, {Morris},
  {Bertram}, {Dov{\v c}iak}, {Duschl}, {Karas}, {K{\"o}nig}, {Krichbaum},
  {Krips}, {Kunneriath}, {Lu}, {Markoff}, {Mauerhan}, {Meyer}, {Moultaka},
  {Mu{\v z}i{\'c}}, {Najarro}, {Pott}, {Schuster}, {Sjouwerman}, {Straubmeier},
  {Thum}, {Vogel}, {Wiesemeyer}, {Zamaninasab}, \& {Zensus}}]{eckart08b}
{Eckart}, A., {Sch{\"o}del}, R., {Garc{\'{\i}}a-Mar{\'{\i}}n}, M., {et~al.}
  2008{\natexlab{b}}, A\&A, 492, 337

\bibitem[{{Eckart} {et~al.}(2006{\natexlab{b}}){Eckart}, {Sch{\"o}del},
  {Meyer}, {Trippe}, {Ott}, \& {Genzel}}]{eckart06b}
{Eckart}, A., {Sch{\"o}del}, R., {Meyer}, L., {et~al.} 2006{\natexlab{b}},
  A\&A, 455, 1

\bibitem[{{Eisenhauer} {et~al.}(2005){Eisenhauer}, {Genzel}, {Alexander},
  {Abuter}, {Paumard}, {Ott}, {Gilbert}, {Gillessen}, {Horrobin}, {Trippe},
  {Bonnet}, {Dumas}, {Hubin}, {Kaufer}, {Kissler-Patig}, {Monnet},
  {Str{\"o}bele}, {Szeifert}, {Eckart}, {Sch{\"o}del}, \&
  {Zucker}}]{eisenhauer05}
{Eisenhauer}, F., {Genzel}, R., {Alexander}, T., {et~al.} 2005, ApJ, 628, 246

\bibitem[{{Falcke} {et~al.}(2009){Falcke}, {Markoff}, \& {Bower}}]{falcke09}
{Falcke}, H., {Markoff}, S., \& {Bower}, G.~C. 2009, \aap, 496, 77

\bibitem[{{Gehrels}(1986)}]{gehrels86}
{Gehrels}, N. 1986, \apj, 303, 336

\bibitem[{{Genzel} {et~al.}(2003){Genzel}, {Sch{\"o}del}, {Ott}, {Eckart},
  {Alexander}, {Lacombe}, {Rouan}, \& {Aschenbach}}]{genzel03}
{Genzel}, R., {Sch{\"o}del}, R., {Ott}, T., {et~al.} 2003, Nature, 425, 934

\bibitem[{{Ghez} {et~al.}(2004){Ghez}, {Wright}, {Matthews}, {Thompson}, {Le
  Mignant}, {Tanner}, {Hornstein}, {Morris}, {Becklin}, \& {Soifer}}]{ghez04}
{Ghez}, A., {Wright}, S., {Matthews}, K., {et~al.} 2004, ApJ, 601, L159

\bibitem[{{Ghez} {et~al.}(2005){Ghez}, {Hornstein}, {Lu}, {Bouchez}, {Le
  Mignant}, {van Dam}, {Wizinowich}, {Matthews}, {Morris}, {Becklin},
  {Campbell}, {Chin}, {Hartman}, {Johansson}, {Lafon}, {Stomski}, \&
  {Summers}}]{ghez05b}
{Ghez}, A.~M., {Hornstein}, S.~D., {Lu}, J.~R., {et~al.} 2005, ApJ, 635, 1087

\bibitem[{{Ghez} {et~al.}(2008){Ghez}, {Salim}, {Weinberg}, {Lu}, {Do}, {Dunn},
  {Matthews}, {Morris}, {Yelda}, {Becklin}, {Kremenek}, {Milosavljevic}, \&
  {Naiman}}]{ghez08}
{Ghez}, A.~M., {Salim}, S., {Weinberg}, N.~N., {et~al.} 2008, ApJ, 689, 1044

\bibitem[{{Gillessen} {et~al.}(2006){Gillessen}, {Eisenhauer}, {Quataert},
  {Genzel}, {Paumard}, {Trippe}, {Ott}, {Abuter}, {Eckart}, {Lagage},
  {Lehnert}, {Tacconi}, \& {Martins}}]{gillessen06}
{Gillessen}, S., {Eisenhauer}, F., {Quataert}, E., {et~al.} 2006, ApJ, 640,
  L163

\bibitem[{{Gillessen} {et~al.}(2009){Gillessen}, {Eisenhauer}, {Trippe},
  {Alexander}, {Genzel}, {Martins}, \& {Ott}}]{gillessen09}
{Gillessen}, S., {Eisenhauer}, F., {Trippe}, S., {et~al.} 2009, \apj, 692, 1075

\bibitem[{{Goldwurm} {et~al.}(2003){Goldwurm}, {Brion}, {Goldoni}, {Ferrando},
  {Daigne}, {Decourchelle}, {Warwick}, \& {Predehl}}]{goldwurm03}
{Goldwurm}, A., {Brion}, E., {Goldoni}, P., {et~al.} 2003, ApJ, 584, 751

\bibitem[{{Gould}(1979)}]{gould79}
{Gould}, R.~J. 1979, \aap, 76, 306

\bibitem[{{G{\"u}sten} {et~al.}(2006){G{\"u}sten}, {Nyman}, {Schilke},
  {Menten}, {Cesarsky}, \& {Booth}}]{gusten06}
{G{\"u}sten}, R., {Nyman}, L.~{\AA}., {Schilke}, P., {et~al.} 2006, \aap, 454,
  L13

\bibitem[{{Herrnstein} {et~al.}(2004){Herrnstein}, {Zhao}, {Bower}, \&
  {Goss}}]{herrnstein04}
{Herrnstein}, R.~M., {Zhao}, J.-H., {Bower}, G.~C., \& {Goss}, W.~M. 2004, AJ,
  127, 3399

\bibitem[{{Hornstein} {et~al.}(2007){Hornstein}, {Matthews}, {Ghez}, {Lu},
  {Morris}, {Becklin}, {Rafelski}, \& {Baganoff}}]{hornstein07}
{Hornstein}, S.~D., {Matthews}, K., {Ghez}, A.~M., {et~al.} 2007, ApJ, 667, 900

\bibitem[{{Jansen} {et~al.}(2001){Jansen}, {Lumb}, {Altieri}, {Clavel}, {Ehle},
  {Erd}, {Gabriel}, {Guainazzi}, {Gondoin}, {Much}, {Munoz}, {Santos},
  {Schartel}, {Texier}, \& {Vacanti}}]{jansen01}
{Jansen}, F., {Lumb}, D., {Altieri}, B., {et~al.} 2001, A\&A, 365, L1

\bibitem[{{Kirsch}(2005)}]{kirsch05}
{Kirsch}, M. 2005, {XMM-SOC-CAL-TN-0018: EPIC Status of Calibration and Data
  Analysis}, Tech. rep., {XMM-Newton Science Operations Centre}

\bibitem[{{Krabbe} {et~al.}(2006){Krabbe}, {Iserlohe}, {Larkin}, {Barczys},
  {McElwain}, {Weiss}, {Wright}, \& {Quirrenbach}}]{krabbe06}
{Krabbe}, A., {Iserlohe}, C., {Larkin}, J.~E., {et~al.} 2006, ApJ, 642, L145

\bibitem[{{Krawczynski} {et~al.}(2004){Krawczynski}, {Hughes}, {Horan},
  {Aharonian}, {Aller}, {Aller}, {Boltwood}, {Buckley}, {Coppi}, {Fossati},
  {G{\"o}tting}, {Holder}, {Horns}, {Kurtanidze}, {Marscher}, {Nikolashvili},
  {Remillard}, {Sadun}, \& {Schr{\"o}der}}]{krawczynski04}
{Krawczynski}, H., {Hughes}, S.~B., {Horan}, D., {et~al.} 2004, ApJ, 601, 151

\bibitem[{{Kunneriath} {et~al.}(2008){Kunneriath}, {Eckart}, {Vogel},
  {Sjouwerman}, {Wiesemeyer}, {Sch{\"o}del}, {Baganoff}, {Morris}, {Bertram},
  {Dovciak}, {Dowries}, {Duschl}, {Karas}, {Konig}, {Krichbaum}, {Krips}, {Lu},
  {Markoff}, {Mauerhan}, {Meyer}, {Moultaka}, {Muzic}, {Najarro}, {Schuster},
  {Straubmeier}, {Thum}, {Witzel}, {Zamaninasab}, \& {Zensus}}]{kunneriath08}
{Kunneriath}, D., {Eckart}, A., {Vogel}, S., {et~al.} 2008, Journal of Physics
  Conference Series, 131, 012006

\bibitem[{{Kusunose} \& {Takahara}(2010)}]{kusunose10}
{Kusunose}, M. \& {Takahara}, F. 2010, ArXiv:1011.1712

\bibitem[{{Lenzen} {et~al.}(2003){Lenzen}, {Hartung}, {Brandner}, {Finger},
  {Hubin}, {Lacombe}, {Lagrange}, {Lehnert}, {Moorwood}, \&
  {Mouillet}}]{lenzen03}
{Lenzen}, R., {Hartung}, M., {Brandner}, W., {et~al.} 2003, in Society of
  Photo-Optical Instrumentation Engineers (SPIE) Conference Series, ed.
  M.~{Iye} \& A.~F.~M. {Moorwood}, Vol. 4841, 944--952

\bibitem[{{Li} {et~al.}(2009){Li}, {Shen}, {Miyazaki}, {Huang}, {Sault},
  {Miyoshi}, {Tsuboi}, \& {Tsutsumi}}]{li09}
{Li}, J., {Shen}, Z., {Miyazaki}, A., {et~al.} 2009, \apj, 700, 417

\bibitem[{{Liu} \& {Melia}(2002)}]{liu02}
{Liu}, S. \& {Melia}, F. 2002, ApJ, 566, 77

\bibitem[{{Liu} {et~al.}(2006{\natexlab{a}}){Liu}, {Melia}, \&
  {Petrosian}}]{liu06a}
{Liu}, S., {Melia}, F., \& {Petrosian}, V. 2006{\natexlab{a}}, ApJ, 636, 798

\bibitem[{{Liu} {et~al.}(2004){Liu}, {Petrosian}, \& {Melia}}]{liu04}
{Liu}, S., {Petrosian}, V., \& {Melia}, F. 2004, ApJ, 611, L101

\bibitem[{{Liu} {et~al.}(2006{\natexlab{b}}){Liu}, {Petrosian}, {Melia}, \&
  {Fryer}}]{liu06b}
{Liu}, S., {Petrosian}, V., {Melia}, F., \& {Fryer}, C.~L. 2006{\natexlab{b}},
  ApJ, 648, 1020

\bibitem[{{Longair}(1994)}]{longairII}
{Longair}, M.~S. 1994, {High energy astrophysics. Volume 2. Stars, the Galaxy
  and the interstellar medium.} No. ISBN 0-521-43439-4 (Cambridge, UK:
  {Cambridge University Press})

\bibitem[{{Maitra} {et~al.}(2009){Maitra}, {Markoff}, \& {Falcke}}]{maitra09}
{Maitra}, D., {Markoff}, S., \& {Falcke}, H. 2009, \aap, 508, L13

\bibitem[{{Markoff} {et~al.}(2007){Markoff}, {Bower}, \& {Falcke}}]{markoff07}
{Markoff}, S., {Bower}, G.~C., \& {Falcke}, H. 2007, \mnras, 379, 1519

\bibitem[{{Markoff} {et~al.}(2001){Markoff}, {Falcke}, {Yuan}, \&
  {Biermann}}]{markoff01}
{Markoff}, S., {Falcke}, H., {Yuan}, F., \& {Biermann}, P. 2001, A\&A, 379, L13

\bibitem[{{Marrone} {et~al.}(2008){Marrone}, {Baganoff}, {Morris}, {Moran},
  {Ghez}, {Hornstein}, {Dowell}, {Mu{\~n}oz}, {Bautz}, {Ricker}, {Brandt},
  {Garmire}, {Lu}, {Matthews}, {Zhao}, {Rao}, \& {Bower}}]{marrone08}
{Marrone}, D.~P., {Baganoff}, F.~K., {Morris}, M.~R., {et~al.} 2008, ApJ, 682,
  373

\bibitem[{{Mauerhan} {et~al.}(2005){Mauerhan}, {Morris}, {Walter}, \&
  {Baganoff}}]{mauerhan05}
{Mauerhan}, J.~C., {Morris}, M., {Walter}, F., \& {Baganoff}, F.~K. 2005,
  \apjl, 623, L25

\bibitem[{{Mayer-Hasselwander} {et~al.}(1998){Mayer-Hasselwander}, {Bertsch},
  {Dingus}, {Eckart}, {Esposito}, {Genzel}, {Hartman}, {Hunter}, {Kanbach},
  {Kniffen}, {Lin}, {Michelson}, {Muecke}, {von Montigny}, {Mukherjee},
  {Nolan}, {Pohl}, {Reimer}, {Schneid}, {Sreekumar}, \&
  {Thompson}}]{mayer-hasselwander98}
{Mayer-Hasselwander}, H., {Bertsch}, D., {Dingus}, B., {et~al.} 1998, A\&A,
  335, 161

\bibitem[{Melia(2007)}]{melia07}
Melia, F. 2007, The Galactic Supermassive Black Hole No. ISBN-13:
  978-0691131290 (41 William Street, Princeton, New Jersey 08540: Princeton
  University Press)

\bibitem[{{Melia} \& {Fatuzzo}(2010)}]{melia10}
{Melia}, F. \& {Fatuzzo}, M. 2010, \mnras, L172

\bibitem[{{Meyer} {et~al.}(2006{\natexlab{a}}){Meyer}, {Eckart}, {Sch{\"o}del},
  {Duschl}, {Mu{\v z}i{\'c}}, {Dov{\v c}iak}, \& {Karas}}]{meyer06b}
{Meyer}, L., {Eckart}, A., {Sch{\"o}del}, R., {et~al.} 2006{\natexlab{a}},
  A\&A, 460, 15

\bibitem[{{Meyer} {et~al.}(2007){Meyer}, {Sch{\"o}del}, {Eckart}, {Duschl},
  {Karas}, \& {Dov{\v c}iak}}]{meyer07}
{Meyer}, L., {Sch{\"o}del}, R., {Eckart}, A., {et~al.} 2007, A\&A, 473, 707

\bibitem[{{Meyer} {et~al.}(2006{\natexlab{b}}){Meyer}, {Sch{\"o}del}, {Eckart},
  {Karas}, {Dov{\v c}iak}, \& {Duschl}}]{meyer06a}
{Meyer}, L., {Sch{\"o}del}, R., {Eckart}, A., {et~al.} 2006{\natexlab{b}},
  A\&A, 458, L25

\bibitem[{{Mirabel} {et~al.}(1998){Mirabel}, {Dhawan}, {Chaty}, {Rodriguez},
  {Marti}, {Robinson}, {Swank}, \& {Geballe}}]{mirabel98}
{Mirabel}, I.~F., {Dhawan}, V., {Chaty}, S., {et~al.} 1998, \aap, 330, L9

\bibitem[{{Miyazaki} {et~al.}(2004){Miyazaki}, {Tsutsumi}, \&
  {Tsuboi}}]{miyazaki04}
{Miyazaki}, A., {Tsutsumi}, T., \& {Tsuboi}, M. 2004, \apjl, 611, L97

\bibitem[{{Muno} {et~al.}(2003){Muno}, {Baganoff}, {Bautz}, {Brandt}, {Broos},
  {Feigelson}, {Garmire}, {Morris}, {Ricker}, \& {Townsley}}]{muno03a}
{Muno}, M., {Baganoff}, F., {Bautz}, M., {et~al.} 2003, ApJ, 589, 225

\bibitem[{{Muno} {et~al.}(2005){Muno}, {Pfahl}, {Baganoff}, {Brandt}, {Ghez},
  {Lu}, \& {Morris}}]{muno05a}
{Muno}, M., {Pfahl}, E., {Baganoff}, F., {et~al.} 2005, ApJ, 622, L113

\bibitem[{{Nishiyama} {et~al.}(2009){Nishiyama}, {Tamura}, {Hatano}, {Nagata},
  {Kudo}, {Ishii}, {Sch{\"o}del}, \& {Eckart}}]{nishiyama09}
{Nishiyama}, S., {Tamura}, M., {Hatano}, H., {et~al.} 2009, \apjl, 702, L56

\bibitem[{{Pierce-Price} {et~al.}(2000){Pierce-Price}, {Richer}, {Greaves},
  {Holland}, {Jenness}, {Lasenby}, {White}, {Matthews}, {Ward-Thompson},
  {Dent}, {Zylka}, {Mezger}, {Hasegawa}, {Oka}, {Omont}, \&
  {Gilmore}}]{pierce00}
{Pierce-Price}, D., {Richer}, J.~S., {Greaves}, J.~S., {et~al.} 2000, \apjl,
  545, L121

\bibitem[{{Ponti} {et~al.}(2010){Ponti}, {Terrier}, {Goldwurm}, {Belanger}, \&
  {Trap}}]{ponti10}
{Ponti}, G., {Terrier}, R., {Goldwurm}, A., {Belanger}, G., \& {Trap}, G. 2010,
  ApJ, 714, 732

\bibitem[{{Ponti} {et~al.}(2009){Ponti}, {Trap}, {Goldwurm}, {Ferrando},
  {Terrier}, {B\'elanger}, {Genzel}, {Gillessen}, {Hasinger}, {Dodds-Eden},
  {Predehl}, {Aschenbach}, {Porquet}, {Grosso}, {Clenet}, {Rouan}, {Sakano},
  {Warwick}, {Melia}, {Farhad}, \& {Reid}}]{ponti09}
{Ponti}, G., {Trap}, G., {Goldwurm}, A., {et~al.} 2009, The Astronomer's
  Telegram, 2038, 1

\bibitem[{{Porquet} {et~al.}(2008){Porquet}, {Grosso}, {Predehl}, {Hasinger},
  {Yusef-Zadeh}, {Aschenbach}, {Trap}, {Melia}, {Warwick}, {Goldwurm},
  {B{\'e}langer}, {Tanaka}, {Genzel}, {Dodds-Eden}, {Sakano}, \&
  {Ferrando}}]{porquet08}
{Porquet}, D., {Grosso}, N., {Predehl}, P., {et~al.} 2008, A\&A, 488, 549

\bibitem[{{Porquet} {et~al.}(2003){Porquet}, {Predehl}, {Aschenbach}, {Grosso},
  {Goldwurm}, {Goldoni}, {Warwick}, \& {Decourchelle}}]{porquet03}
{Porquet}, D., {Predehl}, P., {Aschenbach}, B., {et~al.} 2003, A\&A, 407, L17

\bibitem[{{Reid} {et~al.}(2009){Reid}, {Menten}, {Zheng}, {Brunthaler}, \&
  {Xu}}]{reid09}
{Reid}, M.~J., {Menten}, K.~M., {Zheng}, X.~W., {Brunthaler}, A., \& {Xu}, Y.
  2009, \apj, 705, 1548

\bibitem[{{Rousset} {et~al.}(2003){Rousset}, {Lacombe}, {Puget}, {Hubin},
  {Gendron}, {Fusco}, {Arsenault}, {Charton}, {Feautrier}, {Gigan}, {Kern},
  {Lagrange}, {Madec}, {Mouillet}, {Rabaud}, {Rabou}, {Stadler}, \&
  {Zins}}]{rousset03}
{Rousset}, G., {Lacombe}, F., {Puget}, P., {et~al.} 2003, in Society of
  Photo-Optical Instrumentation Engineers (SPIE) Conference Series, ed. P.~L.
  {Wizinowich} \& D.~{Bonaccini}, Vol. 4839, 140--149

\bibitem[{{Sakano} {et~al.}(2004){Sakano}, {Warwick}, {Decourchelle}, \&
  {Predehl}}]{sakano04}
{Sakano}, M., {Warwick}, R., {Decourchelle}, A., \& {Predehl}, P. 2004, MNRAS,
  350, 129

\bibitem[{{Siringo} {et~al.}(2009){Siringo}, {Kreysa}, {Kov{\'a}cs},
  {Schuller}, {Wei{\ss}}, {Esch}, {Gem{\"u}nd}, {Jethava}, {Lundershausen},
  {Colin}, {G{\"u}sten}, {Menten}, {Beelen}, {Bertoldi}, {Beeman}, \&
  {Haller}}]{siringo09}
{Siringo}, G., {Kreysa}, E., {Kov{\'a}cs}, A., {et~al.} 2009, \aap, 497, 945

\bibitem[{{Str{\"u}der} {et~al.}(2001){Str{\"u}der}, {Briel}, {Dennerl},
  {Hartmann}, {Kendziorra}, {Meidinger}, {Pfeffermann}, {Reppin}, {Aschenbach},
  {Bornemann}, {Br{\"a}uninger}, {Burkert}, {Elender}, {Freyberg}, {Haberl},
  {Hartner}, {Heuschmann}, {Hippmann}, {Kastelic}, {Kemmer}, {Kettenring},
  {Kink}, {Krause}, {M{\"u}ller}, {Oppitz}, {Pietsch}, {Popp}, {Predehl},
  {Read}, {Stephan}, {St{\"o}tter}, {Tr{\"u}mper}, {Holl}, {Kemmer}, {Soltau},
  {St{\"o}tter}, {Weber}, {Weichert}, {von Zanthier}, {Carathanassis}, {Lutz},
  {Richter}, {Solc}, {B{\"o}ttcher}, {Kuster}, {Staubert}, {Abbey}, {Holland},
  {Turner}, {Balasini}, {Bignami}, {La Palombara}, {Villa}, {Buttler},
  {Gianini}, {Lain{\'e}}, {Lumb}, \& {Dhez}}]{struder01}
{Str{\"u}der}, L., {Briel}, U., {Dennerl}, K., {et~al.} 2001, A\&A, 365, L18

\bibitem[{{Terrier} {et~al.}(2010){Terrier}, {Ponti}, {B{\'e}langer},
  {Decourchelle}, {Tatischeff}, {Goldwurm}, {Trap}, {Morris}, \&
  {Warwick}}]{terrier10}
{Terrier}, R., {Ponti}, G., {B{\'e}langer}, G., {et~al.} 2010, \apj, 719, 143

\bibitem[{{Trap} {et~al.}(2010){Trap}, {Goldwurm}, {Terrier}, {Dodds-Eden},
  {Gillessen}, {Genzel}, {Pantin}, {Lagage}, {Ferrando}, {B{\'e}langer},
  {Porquet}, {Grosso}, {Yusef-Zadeh}, \& {Melia}}]{trap10}
{Trap}, G., {Goldwurm}, A., {Terrier}, R., {et~al.} 2010, Adv. Space Res., 45,
  507

\bibitem[{{Trippe} {et~al.}(2007){Trippe}, {Paumard}, {Ott}, {Gillessen},
  {Eisenhauer}, {Martins}, \& {Genzel}}]{trippe07}
{Trippe}, S., {Paumard}, T., {Ott}, T., {et~al.} 2007, \mnras, 375, 764

\bibitem[{{Turner} {et~al.}(2001){Turner}, {Abbey}, {Arnaud}, {Balasini},
  {Barbera}, {Belsole}, {Bennie}, {Bernard}, {Bignami}, {Boer}, {Briel},
  {Butler}, {Cara}, {Chabaud}, {Cole}, {Collura}, {Conte}, {Cros}, {Denby},
  {Dhez}, {Di Coco}, {Dowson}, {Ferrando}, {Ghizzardi}, {Gianotti}, {Goodall},
  {Gretton}, {Griffiths}, {Hainaut}, {Hochedez}, {Holland}, {Jourdain},
  {Kendziorra}, {Lagostina}, {Laine}, {La Palombara}, {Lortholary}, {Lumb},
  {Marty}, {Molendi}, {Pigot}, {Poindron}, {Pounds}, {Reeves}, {Reppin},
  {Rothenflug}, {Salvetat}, {Sauvageot}, {Schmitt}, {Sembay}, {Short},
  {Spragg}, {Stephen}, {Str{\"u}der}, {Tiengo}, {Trifoglio}, {Tr{\"u}mper},
  {Vercellone}, {Vigroux}, {Villa}, {Ward}, {Whitehead}, \& {Zonca}}]{turner01}
{Turner}, M.~J.~L., {Abbey}, A., {Arnaud}, M., {et~al.} 2001, A\&A, 365, L27

\bibitem[{{van der Laan}(1966)}]{vanderlaan66}
{van der Laan}, H. 1966, \nat, 211, 1131

\bibitem[{{Viehmann} {et~al.}(2005){Viehmann}, {Eckart}, {Sch{\"o}del},
  {Moultaka}, {Straubmeier}, \& {Pott}}]{viehmann05}
{Viehmann}, T., {Eckart}, A., {Sch{\"o}del}, R., {et~al.} 2005, A\&A, 433, 117

\bibitem[{{Wang} {et~al.}(2006){Wang}, {Lu}, \& {Gotthelf}}]{wang06}
{Wang}, Q.~D., {Lu}, F.~J., \& {Gotthelf}, E.~V. 2006, MNRAS, 367, 937

\bibitem[{{Wright} \& {Backer}(1993)}]{wright93}
{Wright}, M.~C.~H. \& {Backer}, D.~C. 1993, \apj, 417, 560

\bibitem[{{Yuan} {et~al.}(2003){Yuan}, {Quataert}, \& {Narayan}}]{yuan03}
{Yuan}, F., {Quataert}, E., \& {Narayan}, R. 2003, ApJ, 598, 301

\bibitem[{{Yuan} {et~al.}(2004){Yuan}, {Quataert}, \& {Narayan}}]{yuan04}
{Yuan}, F., {Quataert}, E., \& {Narayan}, R. 2004, ApJ, 606, 894

\bibitem[{{Yusef-Zadeh} {et~al.}(2006{\natexlab{a}}){Yusef-Zadeh}, {Bushouse},
  {Dowell}, {Wardle}, {Roberts}, {Heinke}, {Bower}, {Vila-Vilar{\'o}},
  {Shapiro}, {Goldwurm}, \& {B{\'e}langer}}]{yusef-zadeh06a}
{Yusef-Zadeh}, F., {Bushouse}, H., {Dowell}, C.~D., {et~al.}
  2006{\natexlab{a}}, ApJ, 644, 198

\bibitem[{{Yusef-Zadeh} {et~al.}(2009){Yusef-Zadeh}, {Bushouse}, {Wardle},
  {Heinke}, {Roberts}, {Dowell}, {Brunthaler}, {Reid}, {Martin}, {Marrone},
  {Porquet}, {Grosso}, {Dodds-Eden}, {Bower}, {Wiesemeyer}, {Miyazaki}, {Pal},
  {Gillessen}, {Goldwurm}, {Trap}, \& {Maness}}]{yusef-zadeh09}
{Yusef-Zadeh}, F., {Bushouse}, H., {Wardle}, M., {et~al.} 2009, \apj, 706, 348

\bibitem[{{Yusef-Zadeh} {et~al.}(2006{\natexlab{b}}){Yusef-Zadeh}, {Roberts},
  {Wardle}, {Heinke}, \& {Bower}}]{yusef-zadeh06b}
{Yusef-Zadeh}, F., {Roberts}, D., {Wardle}, M., {Heinke}, C.~O., \& {Bower},
  G.~C. 2006{\natexlab{b}}, ApJ, 650, 189

\bibitem[{{Yusef-Zadeh} {et~al.}(2007){Yusef-Zadeh}, {Wardle}, {Cotton},
  {Heinke}, \& {Roberts}}]{yusef-zadeh07}
{Yusef-Zadeh}, F., {Wardle}, M., {Cotton}, W.~D., {Heinke}, C.~O., \&
  {Roberts}, D.~A. 2007, \apjl, 668, L47

\bibitem[{{Yusef-Zadeh} {et~al.}(2008){Yusef-Zadeh}, {Wardle}, {Heinke},
  {Dowell}, {Roberts}, {Baganoff}, \& {Cotton}}]{yusef-zadeh08}
{Yusef-Zadeh}, F., {Wardle}, M., {Heinke}, C., {et~al.} 2008, ApJ, 682, 361

\bibitem[{{Zhao} {et~al.}(2003){Zhao}, {Young}, {Herrnstein}, {Ho}, {Tsutsumi},
  {Lo}, {Goss}, \& {Bower}}]{zhao03}
{Zhao}, J.-H., {Young}, K.~H., {Herrnstein}, R.~M., {et~al.} 2003, ApJ, 586,
  L29

\bibitem[{{Zylka} \& {Mezger}(1988)}]{zylka88}
{Zylka}, R. \& {Mezger}, P.~G. 1988, \aap, 190, L25

\end{thebibliography}

\end{document}